\documentclass[a4paper,11pt]{article}
\pdfoutput=1 

\usepackage{jheparxiv} 

\usepackage[T1]{fontenc} 

\usepackage{xcolor}
\usepackage{tensor}
\usepackage{amsmath}
\usepackage{amssymb}
\usepackage{mathrsfs}
\usepackage{graphicx}
\usepackage{stmaryrd}
\usepackage{twistor}
\usepackage{commath}
\usepackage{xcolor}
\usepackage{mathtools}

\newcommand{\msf}[1]{\mathsf{#1}}

\title{From AdS correlators to Carrollian amplitudes with the scattering equations}

\author[a]{Tim Adamo,}
\author[a]{Iustin Surubaru}
\author[a,b]{\& Bin Zhu}

\affiliation[a]{School of Mathematics and Maxwell Institute for Mathematical Sciences \\
University of Edinburgh, EH9 3FD, U.K.}

\affiliation[b]{School of Physics, Nankai University, Weijin Road 94, Tianjin 300071, P.R. China}

\emailAdd{t.adamo@ed.ac.uk}
\emailAdd{iustin.surubaru@ed.ac.uk} 
\emailAdd{bzhu@nankai.edu.cn}

\begin{document} 

\abstract{The scattering equations relate massless scattering kinematics to marked points on a Riemann sphere, and underpin remarkable formulae for the full tree-level S-matrices of many interesting QFTs, including cubic biadjoint scalars, Yang-Mills theory and general relativity. The scattering equations arise from worldsheet correlators of ambitwistor string theories, which has enabled their generalisation to anti-de Sitter (AdS) space in certain cases. In this paper, we use the scattering equations and ambitwistor strings to prove the correspondence between an appropriate flat limit of boundary correlators in AdS and Carrollian scattering amplitudes -- massless amplitudes written in position space on the null conformal boundary -- for any number of external states and spacetime dimensions in tree-level, cubic scalar theories. We first derive the Carrollian version of the scattering equations in Minkowski space and their associated Carrollian amplitude formulae, by direct Fourier transform from momentum space and from ambitwistor strings with a Carrollian basis of vertex operators. We then take the flat limit of known formulae for all tree-level boundary correlators of cubic scalar theories in AdS, recovering the Carrollian amplitudes in flat space. In the special case of AdS$_3$, we also make some comments on the flat space limit of spinning boundary correlators.}

\maketitle
\flushbottom

\section{Introduction}

The S-matrix describes the evolution of asymptotic in-states to asymptotic out-states; as such, it is a tautologically holographic quantity encoding the dynamics of any quantum field theory in flat spacetime. Attempts at extending the holographic principle to asymptotically flat spacetimes have often focused on scattering amplitudes as the key quantities for building `bottom-up' approaches. Typically, scattering amplitudes are expressed in a momentum eigenstate basis, but in recent years it has been shown that other bases are potentially more useful in practically exposing the holographic nature of the S-matrix.

In the particular context of massless scattering, this has been investigated under the related banners of celestial and Carrollian holography. Celestial amplitudes are written in terms of a boost eigenstate basis, ensuring that they behave under Lorentz transformations as conformal correlators on the celestial sphere. This alternative basis exposes a rich alternative perspective on the infrared and ultraviolet properties of massless scattering which has been studied in great detail (cf., \cite{Strominger:2017zoo,Raclariu:2021zjz,Pasterski:2021raf,Donnay:2023mrd}). Carrollian amplitudes, on the other hand, are computed in a position space basis at null infinity, producing scattering amplitudes which behave as conformal Carrollian correlators~\cite{Donnay:2022aba,Bagchi:2022emh,Donnay:2022wvx,Bagchi:2025vri,Nguyen:2025zhg}. The two perspectives (celestial and Carrollian) are linked by a simple integral transform.

Of course, to extract meaningful lessons about asymptotically flat holography, it is natural to look for a connection with the best understood incarnation of the holographic principle: the AdS/CFT correspondence~\cite{Maldacena:1997re,Witten:1998qj,Gubser:1998bc}. While it has been known that flat limits of boundary correlators in AdS should produce scattering amplitudes since the earliest days of AdS/CFT (e.g., \cite{Susskind:1998vk,Polchinski:1999ry,Giddings:1999jq,Giddings:1999qu}), formulating a precise prescription to do so is not trivial. In particular, the flat space limit is often singular in some sense: typical coordinate systems used in AdS position space calculations (such as Poincar\'e coordinates) are not well adapted to the flat limit, and the flat limit of AdS momentum space correlators corresponds to certain poles~\cite{Maldacena:2002vr,Raju:2012zr,Maldacena:2011nz}. Yet these difficulties have not prevented a fruitful interchange between amplitudes and AdS/CFT methods (e.g., \cite{Fitzpatrick:2011dm,Paulos:2016fap,Alday:2017vkk,Hijano:2019qmi,Komatsu:2020sag,vanRees:2022zmr}): a particularly good example of this is provided by writing AdS correlators in Mellin variables, which mimic flat space Mandelstam invariants and which encode the flat limit through their `high-energy' regime~\cite{Penedones:2010ue,Fitzpatrick:2011ia,Paulos:2011ie}.

\medskip

Recently, the connection between flat limits of scalar AdS correlators and scattering amplitudes has been precisely realized in the framework of Carrollian amplitudes~\cite{Bagchi:2023cen,Alday:2024yyj}. Here, the key step is to evaluate AdS correlators -- or, equivalently, the constituent ingredients of Witten diagrams -- in an AdS-Bondi coordinate system that smoothly passes to planar Bondi coordinates on Minkowski space in the flat limit. This ensures that, diagram-by-diagram, Witten diagrams pass over to position space Feynman diagrams in the flat limit, giving a correspondence between AdS correlators and scattering amplitudes which holds in the same parametrization (i.e., position space) on \emph{both} sides (unlike the flat limit in Mellin variables, for instance). As expected, there is information loss in this flat limit, with AdS masses washing out in overall factors or derivative operators acting on the resulting massless Carrollian scalar amplitude\footnote{This is by no means the unique way to implement the flat space limit. For instance, it should be possible to obtain massive amplitudes in the flat limit~\cite{Berenstein:2025tts}, although any Carrollian interpretation of such amplitudes is less clear.}.

The fact that this flat limit correspondence holds at the level of contact interactions, bulk-to-bulk and bulk-to-boundary propagators essentially implies that it holds for all scalars correlators/Carrollian amplitudes. However, explicit tests of the flat limit have been restricted to low multiplicity examples (2-, 3- and 4-point functions) due to the complications of computing Carrollian amplitudes for higher numbers of external points~\cite{Donnay:2022aba,Bagchi:2022emh,Donnay:2022wvx,Bagchi:2023fbj,Bagchi:2023cen,Alday:2024yyj,Salzer:2023jqv,Saha:2023abr,Nguyen:2023vfz,Nguyen:2023miw,Mason:2023mti,Liu:2024nfc,Have:2024dff,Stieberger:2024shv,Adamo:2024mqn,Banerjee:2024hvb,Ruzziconi:2024zkr,Ruzziconi:2024kzo,Chakrabortty:2024bvm,Nguyen:2025sqk,Lipstein:2025jfj,deGioia:2025mwt}. Furthermore, the vast majority of studies of Carrollian amplitudes have been restricted to four spacetime dimensions, and the few exceptions in $D=3$~\cite{Surubaru:2025fmg} or $D>4$~\cite{Liu:2024llk,Kulkarni:2025qcx} are still for low numbers of external points.

\medskip

In this paper, we provide an explicit proof of the flat limit correspondence between AdS boundary correlators and Carrollian amplitudes for cubic scalar theories, at tree-level with \emph{any} number of external states and in \emph{arbitrary} spacetime dimension. The key tools which enable this are the \emph{scattering equations}, a set of kinematic constraints which relate massless momenta to the moduli of an auxiliary Riemann sphere with marked points. These scattering equations underpin the famous Cachazo-He-Yuan (CHY) representations of the full tree-level S-matrices of a wide variety of massless QFTs in any number of spacetime dimensions~\cite{Cachazo:2013hca,Cachazo:2014xea}. A remarkable feature of the CHY formulae is universality: different theories correspond to simple replacements of an integrand, which is localised on solutions of the scattering equations. 

By direct Fourier transform, we obtain CHY formulae for Carrollian amplitudes, where the Carrollian `scattering equations' become differential operators containing Carroll boosts acting on Carrollian scalar contact diagrams. These Carrollian CHY formulae hold for every QFT for which a CHY representation is known, and we provide an independent check on the results by deriving them from worldsheet correlators in ambitwistor string theories~\cite{Mason:2013sva}, which are chiral string theories known to produce the CHY formulae in momentum space.

For biadjoint scalar theory (with or without a mass), there is also a CHY-like formula for all tree-level boundary correlation functions in AdS of any dimension~\cite{Eberhardt:2020ewh,Roehrig:2020kck}. The AdS `scattering equations' are also differential operators, encoding the action of boundary conformal generators acting on contact Witten diagrams, suggesting a potential connection with their Carrollian counterparts in flat space. By working in the appropriate AdS-Bondi coordinates, we show that the flat space limit of these scalar AdS CHY formulae is directly related to the Carrollian CHY formulae, providing a proof of the flat limit correspondence at arbitrary multiplicity and in any spacetime dimension.

More specifically, let $\mathcal{C}_{n}(\Delta;\ell)$ be the tree-level, $n$-point boundary correlation function of a cubic scalar theory in AdS$_{D}$ of radius $\ell$, where $\Delta$ encodes the mass of the scalar via $m^2=\ell^{-2}\Delta(\Delta-D+1)$. We prove that
\be\label{FlatLimit*}
\lim_{\ell\to\infty}\ell^{n\left(\frac{D-2}{2}-\Delta\right)}\,\mathcal{C}_{n}(\Delta;\ell)\propto \prod_{i=1}^{n}\partial_{u_i}^{\Delta-1} C_{n}\,,
\ee
where $C_n$ is the tree-level, $n$-point massless Carrollian amplitude (defined to be the Fourier transform of the momentum space amplitude with respect to the frequencies of the external particles) of the cubic scalar theory in flat space, with external states parametrized by points $(u_i,\mathbf{x}_i)\in\R\times S^{D-2}$ on the null conformal boundary.

While there are no such general CHY-like formulae for spinning AdS correlators, in AdS$_3$ there \emph{are} formulae for certain gluon and graviton boundary correlators~\cite{Roehrig:2020kck}. The flat limit of spinning AdS correlators is notoriously more subtle than for scalars (cf., \cite{Li:2021snj,Berenstein:2025qhb}), but we are nevertheless able to make some interesting statements about the flat limit here as well.

\medskip

The paper is organized as follows. In section \ref{sec:2}, after reviewing some basics of Carrollian amplitudes and CHY formulae, we derive Carrollian scattering equations and the Carrollian CHY formulae for in generic spacetime dimensions. We show that these formulae also arise from ambitwistor strings when the BRST cohomology is expressed in a Carrollian basis. In section \ref{sec:3}, we review the basic elements of scattering equations in AdS introduced by \cite{Eberhardt:2020ewh,Roehrig:2020kck} for scalar boundary correlators. In section \ref{sec:4}, we show how the flat space limit of the AdS CHY formulae is naturally implemented in Bondi coordinates, leading to a matching to the Carrollian CHY formulae for biadjoint scalar shown in section \ref{sec:2}. In section \ref{sec:5}, we comment on the spinning case in AdS$_3$ and its flat space limit. Section \ref{sec:6} concludes with some future directions.


\section{Carrollian scattering equations and amplitudes} \label{sec:2}

The complete tree-level S-matrices of a wide variety of massless QFTs are captured by the CHY formulae~\cite{Cachazo:2013hca,Cachazo:2013iea,Cachazo:2014nsa,Cachazo:2014xea}, which present scattering amplitudes in terms of localized integrals over the moduli space of a punctured Riemann sphere. This localization is accomplished by the \emph{scattering equations}, which fix the moduli of the punctured Riemann sphere in terms of the on-shell momenta of the external particles in the scattering process. The scattering equations and CHY formulae have a worldsheet origin, arising from genus zero correlation functions in a theory governing holomorphic maps from a worldsheet to the space of scale-less, complexified null geodesics in Minkowski space, known as \emph{ambitwistor strings}~\cite{Mason:2013sva,Berkovits:2013xba,Adamo:2013tsa}.

While the scattering equations and CHY formulae are typically formulated for massless scattering amplitudes in a momentum eigenstate basis, there is no obstruction to generalising them to amplitudes computed in other bases of massless external states. Indeed, this has already been done for scattering amplitudes computed in the `conformal primary basis'~\cite{Pasterski:2016qvg,Pasterski:2017kqt} of external states~\cite{Adamo:2019ipt,Casali:2020uvr}, in which case the scattering equations of momentum space are replaced by certain differential operators acting with respect to the scaling dimensions of the external states.

In this section, we define the scattering equations for tree-level \emph{Carrollian amplitudes}, massless scattering amplitudes computed in a position-space basis on the null conformal boundary of Minkowski space. After reviewing the basic definition of these Carrollian amplitudes, we derive the Carrollian scattering equations (really differential operators acting with respect to the locations of the external states in Bondi time) using two methods. First, we perform a Fourier transform to go from the standard momentum space basis to the Carrollian one. We then consider ambitwistor strings with vertex operators expressed in the Carrollian basis, computing their correlation functions. The two procedures produce identical Carrollian scattering equations and CHY formulae (as expected), which can be defined for a wide variety of massless QFTs in any number of spacetime dimensions.


\subsection{Carrollian scattering amplitudes}

Carrollian amplitudes are, roughly speaking, massless scattering amplitudes computed in a position space, rather than momentum space, basis. One can naturally ask what is meant by the `positions' of external massless particles which are, by definition, located in the infinite past and future for a scattering process. To make this notion precise, one naturally works with a conformal compactification of Minkowski spacetime, where massless outgoing/incoming massless particles can be associated with locations on $\scri^{\pm}$, the future/past null conformal boundaries. 

To capture both incoming and outgoing states simultaneously, it is useful to work with \emph{planar Bondi coordinates}~\cite{Donnay:2022wvx} which cover the entire conformal compactification (rather than the more standard retarded/advanced Bondi coordinates, which include only one of $\scri^{\pm}$)\footnote{It should be noted that such a global treatment will \emph{not} be possible for generic asymptotically flat spacetimes, and is closely tied to the analyticity of Minkowski space.}. Consider $D=d+2$-dimensional Minkowski spacetime $\mathbb{R}^{d+1,1}$; the planar Bondi coordinates $(u,r,\mathbf{x})$ are defined for $u,r\in\R$ and $\mathbf{x}\in S^d$, and can be related to standard Cartesian Minkowski coordinates by
\begin{equation}
    X^{\mu} = \frac{u\, \Box_\mathbf{x}q^{\mu}}{d}+r\, q^{\mu} = \left( u+ \frac{r}{2}\,(1+|\mathbf{x}|^2),\, r\, \mathbf{x},\,-u-\frac{r}{2}\,(|\mathbf{x}|^2-1)\right) \, ,
\end{equation}
where $|\mathbf{x}|^2$ denotes the inner product on the unit $d$-sphere and the null vector
\begin{equation}
    q^{\mu}(\mathbf{x}) = \frac{1}{2}\left(1+|\mathbf{x}|^2,\, 2\mathbf{x},\, 1-|\mathbf{x}|^2\right) \, , \label{eq:defqmu}
\end{equation}
corresponds to a point $\mathbf{x}$ on the celestial $d$-sphere.

The Minkowski metric in these planar Bondi coordinates is
\begin{equation}
    \d s^2_{\R^{d+1,1}}=-2\,\d u\,\d r+r^2\,|\d\mathbf{x}|^2 \, , \label{eq:gflatBondi}
\end{equation}
with conformal compactification achieved by conformal factor $R:=r^{-1}$, leaving
\be\label{CCompact}
\d s^{2}_{\R^{d+1,1}}\rightarrow R^{2}\,\d s^2_{\R^{d+1,1}}=2\,\d u\,\d R+|\d\mathbf{x}|^2\,,
\ee
for $R\in(-\infty,0^{-})\cup(0^+,+\infty)$. The conformal boundaries $\scri^{\pm}$ now correspond to  $R\to 0^{\pm}$, both resulting in the same degenerate null metric
\begin{equation}
    \d s^2_{\scri^{\pm}} = 0\times \d u^2+ |\d\mathbf{x}|^2 \, , \label{eq:gninf}
\end{equation}
on $\scri^{\pm}\cong\R\times S^{d}$. In addition to covering both past and future null infinity in the conformal compactification, these planar Bondi coordinates have the added advantage that they emerge naturally from the flat limit of Bondi coordinates on AdS, as we will see later.

\medskip

Working in these planar Bondi coordinates, \emph{Carrollian amplitudes} (cf., \cite{Bagchi:2022emh,Donnay:2022aba,Donnay:2022wvx,Mason:2023mti,Salzer:2023jqv,Nguyen:2023miw}) are scattering amplitudes of massless particles expressed in a basis where each external state is associated with a point on the conformal boundary $\scri^-\cup\scri^+$; in essence these are massless scattering amplitudes written in position space, rather than momentum space. This change of basis can be directly implemented by Fourier transforming standard momentum space amplitudes with respect to the frequency of the external massless particles.

The momentum vector of a massless particle can be parametrized as
\begin{equation}\label{nullmompar}
    k^\mu=\alpha\,\omega\,  q^\mu\,,
\end{equation}
where $\alpha=\pm1$ indicates whether the particle is outgoing/incoming, $\omega$ is the frequency (or light-cone energy) and $q^{\mu}$ is a null vector of the form \eqref{eq:defqmu}, defined by a point on the celestial $d$-sphere. Consider the scattering of $n$ massless particles, each parametrized by $\{\omega_i,q_i,\alpha_i\}$, for $i=1,\ldots,n$, as well as a spin $J_i$ and other quantum numbers (e.g., charge, colour) which will be suppressed in our notation unless necessary. The corresponding momentum space scattering amplitude will be denoted by
\be\label{momspaceamp}
A_{n}(\{\boldsymbol{\omega},\mathbf{q}\}^{\boldsymbol{\alpha}}_{\mathbf{J}})\,,
\ee
with
\be\label{boldnotation}
\{\boldsymbol{\omega},\mathbf{q}\}^{\boldsymbol{\alpha}}_{\mathbf{J}}\equiv\{\omega_1, q_1\}^{\alpha_1}_{J_1},\ldots,\{\omega_n, q_n\}^{\alpha_n}_{J_n}\,,
\ee
abbreviating the collection of parameters for the external massless particles.

Given such a momentum space amplitude, we define the corresponding Carrollian amplitude to be the Fourier transform with respect to the frequencies of the external particles:
\be\label{eq:nCarrollian}
C_{n}(\{\mathbf{u},\mathbf{q}\}^{\boldsymbol{\alpha}}_{\boldsymbol{J}}):=\prod_{i=1}^{n}\left(\int_{0}^{\infty}\frac{\d\omega_i}{2\pi}\,\e^{\im\alpha_i\,\omega_i\,u_i}\right)A_{n}(\{\boldsymbol{\omega},\mathbf{q}\}^{\boldsymbol{\alpha}}_{\mathbf{J}})\,.
\ee
The result of this Fourier transform\footnote{Note that it is also possible to define Carrollian amplitudes using the modified Mellin transform~\cite{Banerjee:2018fgd,Banerjee:2018gce}
\be\label{modMellin}
\mathcal{C}_{n}(\{\mathbf{u},\mathbf{q},\boldsymbol{\Delta}\}^{\boldsymbol{\alpha}}_{\boldsymbol{J}}):=\prod_{i=1}^{n}\left(\int_{0}^{\infty}\frac{\d\omega_i}{2\pi}\,\omega_i^{\Delta_i-1}\,\e^{\im\alpha_i\,\omega_i\,u_i}\right)A_{n}(\{\boldsymbol{\omega},\mathbf{q}\}^{\boldsymbol{\alpha}}_{\mathbf{J}})\,.
\ee
In recent work on Carrollian amplitudes in general dimension~\cite{Kulkarni:2025qcx}, it was argued that setting $\Delta_i=(D-2)/2$ is a natural choice by relating to the extrapolation of bulk fields. However, we will see that the simpler, dimension-agnostic choice of $\Delta_i=1$ still leads to sensible results upon comparison with flat limits of AdS correlators.} is that each external particle is now labeled by $(u_i,q_i)\in\R\times S^{d}$ -- namely, by a point on the conformal boundary $\scri^{\alpha_i}$.

Equivalently, Carrollian amplitudes can be defined as massless S-matrix elements where the external wavefunctions are expressed not in a momentum eigenstate basis but rather in a basis of $\Delta=1$ Carrollian conformal primary wavefunctions~\cite{Banerjee:2018fgd,Banerjee:2018gce,Pasterski:2021dqe,Bagchi:2022emh,Donnay:2022aba,Donnay:2022wvx}. For a massless scalar, these wavefunctions are given explicitly by 
\begin{equation}
    \phi^\alpha(X;u,q) = \int_0^\infty \frac{\d\omega}{2\pi}\, \e^{\im\,\alpha\,\omega\, u}\, \e^{\mathrm{i}\,\alpha\, \omega\, q \cdot X-\varepsilon\, \omega} = \frac{\alpha\, \mathrm{i}}{2\pi\,(u+q\cdot X +\mathrm{i}\,\alpha\, \varepsilon)}\,,  \label{eq:Fourierpw}
\end{equation}
where $\varepsilon>0$ is a small regulator to ensure convergence of the Fourier integral. As discussed in \cite{Alday:2024yyj}, the wavefunctions \eqref{eq:Fourierpw} act as bulk-to-boundary propagators for massless scalar fields in Minkowski space, written in planar Bondi coordinates.

It is straightforward to define spinning Carrollian conformal primary wave functions. For instance, Carrollian gluon and graviton wavefunctions can be written as
\begin{equation}
    A_{\mu}^{\mathsf{a},\alpha}= \mathsf{T^a}\, \epsilon_\mu\, \phi^\alpha(X;q,u) \, ,
\end{equation}
\begin{equation}
    h_{\mu\nu}^{\alpha} = \epsilon_{(\mu}\, \tilde{\epsilon}_{\nu)}\, \phi^\alpha(X;q,u) \, ,
\end{equation}
where the polarization vectors $\epsilon_\mu$, $\tilde{\epsilon}_\nu$ can be chosen such that $\epsilon\cdot\tilde{\epsilon}=0$, with each only depending upon $q^\mu$~\cite{Donnay:2022wvx}.

 
\subsection{Review: Scattering equations \& CHY formulae}
\label{CHYflat}

The \emph{scattering equations} are a set of algebraic constraints relating massless kinematics to the moduli of a punctured Riemann sphere~\cite{Fairlie:1972zz,Fairlie:2008dg,Cachazo:2013gna}. For a $n$-point scattering process, the scattering equations are
\be\label{momSE}
E_i:=\sum_{j\neq i}\frac{k_i\cdot k_j}{z_{ij}}=0\,,
\ee
for each $i=1,\ldots,n$, where $\{k^{\mu}_i\}$ are the massless momenta of the external particles and $\{z_i\}$ are the locations of $n$ punctures on the Riemann sphere, written in local stereographic coordinates, with $z_{ij}:=z_i-z_j$. These equations are invariant under the action of SL$(2,\C)$ -- or M\"obius transformations -- on the $\{z_i\}$, so in fact only $n-3$ of them are independent. The scattering equations can be re-arranged to give a polynomial system for the $\{z_i\}$~\cite{Dolan:2014ega}, which generically has $(n-3)!$ distinct solutions~\cite{Dolan:2015iln}.

Among their many remarkable properties -- including describing the high-energy saddle points of string theory~\cite{Gross:1987ar} and manifesting KLT orthogonality~\cite{Cachazo:2013gna} -- is the fact that the scattering equations underpin all-multiplicity formulae for tree-level scattering amplitudes of a wide variety of massless field theories, known as the Cachazo-He-Yuan (CHY) formulae~\cite{Cachazo:2013hca,Cachazo:2013iea,Cachazo:2014nsa,Cachazo:2014xea}. The general form of the CHY formulae is
\be\label{eq:CHY1}
A_{n}=\delta^{D}\!\left(\sum_{i=1}^{n}k_i\right) \int\frac{\d^{n}z}{[\mathrm{vol}\,\SL(2,\C)]^2}\,\prod_{j=1}^{n}\delta(E_j)\,\mathcal{I}\,\tilde{\mathcal{I}}\,,
\ee
where the integral is over the locations of the $n$ marked points $\{z_i\}$. The integrands $\mathcal{I}$, $\tilde{\mathcal{I}}$ are rational functions of the external quantum numbers (momenta, polarisations, charges, etc.) and marked points whose choice determines which the theory's amplitudes are being described. The quotient by two copies of $\mathrm{vol}\,\SL(2,\C)$ should be understood in the Faddeev-Popov sense of removing the M\"obius redundancies in the formula. More precisely,
\be\label{SL2quot}
\frac{\d^{n}z}{\mathrm{vol}\,\SL(2,\C)}:=z_{ab}\,z_{bc}\,z_{ca}\,\prod_{i\neq a,b,c}\d z_i\,, \qquad \frac{1}{\mathrm{vol}\,\SL(2,\C)}\prod_{j=1}^{n}\delta(E_j):=z_{ab}\,z_{bc}\,z_{ca}\,\prod_{j\neq a,b,c}\delta(E_j)\,,
\ee
for any choice of punctures $z_a,z_b,z_c$ whose locations are fixed to absorb the SL$(2,\C)$ redundancies.

Consequently, the integral in \eqref{eq:CHY1} is over the $(n-3)$-dimensional moduli space $\cM_{0,n}$ with precisely $n-3$ delta functions. In other words, the integral is completely localised against the scattering equations, and obtaining an explicit expression for the amplitude boils down to an algebraic problem: determining the $(n-3)!$ solutions of the scattering equations.

Explicit integrands $\mathcal{I}$, $\tilde{\mathcal{I}}$ have now been constructed for a wide variety of massless QFTs~\cite{Cachazo:2014xea}, but let us review three of the most prominent examples: 

{\it Cubic biadjoint scalar:}
\begin{equation}\label{PTfact}
    \mathcal{I} = \tilde{\mathcal{I}} = \text{PT}_n := \sum_{\sigma\in S_n\setminus\Z_n}\frac{\text{Tr}(\mathsf{T}^{\mathsf{a}_{\sigma(1)}}\mathsf{T}^{\mathsf{a}_{\sigma(2)}}\cdots \mathsf{T}^{\mathsf{a}_{\sigma(n)}})}{z_{\sigma(1)\sigma(2)}\,z_{\sigma(2)\sigma(3)}\cdots z_{\sigma(n)\sigma(1)}}\,,
\end{equation}
where $\{\mathsf{T}^{\mathsf{a}}\}$ are generators of the adjoint representation of a semi-simple Lie algebra and the sum is over non-cyclically-related permutations on the labels of the external particles. 

{\it Yang-Mills:}
\begin{equation}\label{YMint}
    \mathcal{I} = \text{Pf}'\Psi(k,\epsilon) \, ,  \quad \tilde{\mathcal{I}} = \text{PT}_n \, ,
\end{equation}
where $\text{Pf}'\Psi$ is a certain `reduced Pfaffian.' The $2n\times 2n$ skew-symmetric matrix $\Psi$ has block form
\begin{equation}
    \Psi(k,\epsilon) = \Bigg(
    \begin{matrix}
A & -C^{T} \\
C & B 
\end{matrix} \Bigg)\, ,
\end{equation}
with entries
\begin{equation}
    A_{ij} = \left\{ {\frac{k_i\cdot k_j}{z_{ij}} \quad i\neq j\atop 0 \quad i=j}\right. \,,  \quad B_{ij} = \left\{ {\frac{\epsilon_i\cdot \epsilon_j}{z_{ij}} \quad i\neq j\atop 0 \quad i=j}\right. \, , \quad C_{ij} = \left\{ {\frac{\epsilon_i\cdot k_j}{z_{ij}} \quad i\neq j\atop -\sum_{l\neq i} \frac{\epsilon_i\cdot k_l}{z_{il}} \quad i=j}\right.\,, \label{eq:compoPsi}
\end{equation}
for $\{\epsilon_i^{\mu}\}$ the on-shell polarization vectors of the external gluons. The reduced Pfaffian is then defined by 
\begin{equation}
    \text{Pf}'\Psi := 2\, \frac{(-1)^{i+j}}{z_{ij}}\,\text{Pf}(\Psi_{ij}^{ij}) \, ,
\end{equation}
with $\text{Pf}(\Psi_{ij}^{ij})$ the Pfaffian of $\Psi$ with rows and columns $i,j$ removed. The CHY formula corresponding to \eqref{YMint} is independent of the choice of the rows and columns removed to define $\mathrm{Pf}'\Psi$, as the notation suggests.

{\it NS-NS gravity:}
\begin{equation}\label{NSNSint}
    \mathcal{I} = \text{Pf}'\Psi(k,\epsilon)\, ,  \quad \tilde{\mathcal{I}} = \text{Pf}'\Psi_n (k,\tilde{\epsilon}) \, .
\end{equation}
The second polarization vector ensures that the product of reduced Pfaffians encodes the information of an `NS-NS graviton' for each external state:
\be\label{NSNSpol}
\epsilon_{\mu}\,\tilde{\epsilon}_{\nu}=\underbrace{\epsilon_{(\mu}\,\tilde{\epsilon}_{\nu)}-\eta_{\mu\nu}\,\frac{\epsilon\cdot\tilde{\epsilon}}{D}}_{\text{graviton}}+\underbrace{\epsilon_{[\mu}\,\tilde{\epsilon_{\nu]}}}_{B\text{-field}}+\underbrace{\epsilon\cdot\tilde{\epsilon}\,\frac{\eta_{\mu\nu}}{D}}_{\text{dilaton}}\,.
\ee
To obtain pure graviton amplitudes, one simply chooses $\tilde{\epsilon}_{\mu}=\epsilon_{\mu}$ so that $\epsilon\cdot\tilde{\epsilon}=0$.

These (and other) choices of integrand can be verified as correctly describing the amplitudes of a desired QFT by checking the factorization properties of the resulting formula \eqref{eq:CHY1}~\cite{Dolan:2013isa}.


\subsection{From momentum to position space}

As defined in \eqref{eq:nCarrollian}, Carrollian amplitudes in position space can be obtained via a Fourier transform of the corresponding momentum space amplitudes. This should still be true when the momentum space input is a scattering amplitude expressed in the CHY form \eqref{eq:CHY1}. In particular, $n$-point tree-level Carrollian amplitudes can be obtained as
\begin{multline}\label{eq:CHYF}
    C_n= \int\frac{\d^{n}z}{[\mathrm{vol}\,\SL(2,\C)]^2}\left(\prod_{i=1}^n\int_{0}^{\infty} \frac{\d\omega_i}{2\pi}\, \e^{\mathrm{i}\,\alpha_i\,\omega_i\, u_i} \right)\delta^{D}\!\left(\sum_{j=1}^{n}\alpha_j\,\omega_j\,q_j\right) \\
    \times\, \prod_{k=1}^{n}\delta\!\left(\sum_{l\neq k}\frac{\alpha_{k}\,\alpha_l\,\omega_k\,\omega_l\,q_k\cdot q_l}{z_{kl}}\right)\,\mathcal{I}\,\tilde{\mathcal{I}}\,,
\end{multline}
where we have explicitly implemented the parametrization \eqref{nullmompar} for the external momenta.    

While \eqref{eq:CHYF} is, tautologically, an expression for Carrollian amplitudes, it is not a particularly illuminating one. To proceed, we follow the same strategy as~\cite{Casali:2020uvr} when refining the CHY formulae for celestial amplitudes~\cite{Adamo:2019ipt}. Observe that the momentum operator $\mathbb{P}_\mu$ acts on Carrollian amplitudes in a simple way:
\begin{equation}\label{momact}
    \P_{\mu}\, \int_0^{\infty} \d\omega\, \e^{\mathrm{i} \alpha\, \omega\, u}\, A(k) = -\mathrm{i}\,q_\mu\, \partial_u \int_0^{\infty}\d\omega \e^{\mathrm{i} \alpha\, \omega\, u}\, A(k) \, .
\end{equation}
Therefore, the action of the translation symmetry generators on Carrollian amplitudes can be defined in terms of the differential operators
\begin{equation}\label{Kdef}
    \mathbf{K}_\mu := -\mathrm{i}\,q_\mu\, \partial_u \, .
\end{equation}
In contrast to celestial amplitudes, where the momentum operator shifts the conformal dimensions of celestial conformal primary wavefunction~\cite{Casali:2020uvr}, for Carrollian amplitudes the momentum operator is simply a partial $u$-derivative.

Poincar\'e invariance then dictates that Carrollian amplitudes are annihilated by
\begin{equation}\label{Poincare}
    \sum_{i=1}^n \mathbf{K}_i^\mu = -\im\,\sum_{i=1}^n q^\mu_i\,\partial_{u_i} \, .
\end{equation}
To refine the Carrollian CHY formula, one can simply make the replacements
\begin{equation}
    \omega_i \rightarrow -\mathrm{i}\,\partial_u \, , \quad k^{\mu}_i \rightarrow\mathbf{K}^{\mu}_i\,,
\end{equation}
inside the scattering equations and integrands $\mathcal{I}$, $\tilde{\mathcal{I}}$. Now, the Carrollian amplitude \eqref{eq:CHYF} can be written as
\begin{equation}
 C_n= \int\frac{\d^{n}z}{[\mathrm{vol}\,\SL(2,\C)]^2}\,\prod_{i=1}^{n}\delta(\mathbf{E}_i)\,\mathcal{I}\,\tilde{\mathcal{I}}\,S_{n}\,. \label{eq:CarrollianCHY}
\end{equation}
Here, the Carrollian scattering equations are in fact differential operators
\begin{equation}\label{CarrSE}
    \mathbf{E}_i := \sum_{j\neq i} \frac{\mathbf{K}_i\cdot \mathbf{K}_j }{z_{ij}}\, ,
\end{equation}
for
\begin{equation}
    \mathbf{K}_i \cdot \mathbf{K}_j = - q_i\cdot q_j\, \partial_{u_i} \partial_{u_j} \, , \label{eq:KidotKj}
\end{equation}
and the integrands $\mathcal{I}$, $\tilde{\mathcal{I}}$ will, generically, be differential operators themselves. For instance, the matrix $\Psi$ appearing in the integrands for Yang-Mills theory \eqref{YMint} and NS-NS gravity \eqref{NSNSint} has several entries which are promoted to differential operators:
\be\label{PsiOps}
A_{ij}=\frac{\mathbf{K}_i\cdot\mathbf{K}_j}{z_{ij}}\,, \qquad C_{ij}=\frac{\epsilon_{i}\cdot\mathbf{K}_j}{z_{ij}}\,, \qquad C_{ii}=-\sum_{j\neq i}\frac{\epsilon_i\cdot\mathbf{K}_j}{z_{ij}}\,.
\ee
All of these differential operators now act on the quantity
\begin{equation}
    S_n:=\left(\prod_{i=1}^{n}\int_0^{\infty} \frac{\d\omega_i}{2\pi}\, \e^{\mathrm{i}\alpha_i\, \omega_i\, u_i}\right) \delta^D\!\left(\sum_{j=1}^n \alpha_j\, \omega_j q_j\right)  \, . \label{eq:Sn}
\end{equation}
Below, we will show that this object is equal to the $n$-point Carrollian contact diagram, and evaluate it explicitly.

However, before proceeding, one may rightly be concerned about what is meant by delta functions of differential operators, $\delta(\mathbf{E}_i)$, appearing in the Carrollian CHY formula \eqref{eq:CarrollianCHY}. The technical meaning of such an expression is defined in terms of the integral representation of the delta function itself:
\be\label{deltaSEq1}
\delta(\mathbf{E}_i)=\int_{-\infty}^{\infty}\frac{\d r_i}{2\pi}\,\e^{\im\,r_i\,\mathbf{E}_i}\,.
\ee
The formal meaning of the expression on the right-hand side of this equation \emph{is} clear: one acts with the (infinite-order) differential operator
\be\label{deltaSEq2}
\e^{\im\,r_i\,\mathbf{E}_i}=\sum_{a=0}^{\infty}\frac{(\im\,r_i)^a}{a!}\,\mathbf{E}_i^{a}\,,
\ee
and then performs the integral\footnote{If required, convergence of this integral can be insured by introducing a regulator $-r_i\,\varepsilon_i$ into the argument of the exponential, for $0<\varepsilon_i\ll1$.} over the parameter $r_i$. 

Hence, the Carrollian CHY formula is more accurately expressed as
\be\label{CarCHY2}
C_n= \int\frac{\d^{n}z}{[\mathrm{vol}\,\SL(2,\C)]^2}\,\left(\prod_{i=1}^{n}\int_{-\infty}^{\infty}\frac{\d r_i}{2\pi}\,\e^{\im\,r_i\,\mathbf{E}_i}\right)\,\mathcal{I}\,\tilde{\mathcal{I}}\,S_{n}\,,
\ee
in terms of the exponentiated differential operators $\{\mathbf{E}_i\}$. We will continue to abuse terminology by referring to the operator \eqref{deltaSEq1} as a Carrollian scattering equation, even though this is not an algebraic equation (as in momentum space), but rather a certain integro-differential operator.


\subsection*{Carrollian contact diagrams}

We conclude our exposition of the Carrollian scattering equations and CHY formulae by taking a closer look at the universal object $S_n$, given by \eqref{eq:Sn}, which appears in the amplitude formulae. Observe that
\begin{equation} 
\begin{split}
    S_n &=\left(\prod_{i=1}^{n}\int_0^{\infty} \frac{\d\omega_i}{2\pi}\, \e^{\mathrm{i}\alpha_i\, \omega_i\, u_i}\right)\int_{\R^{d+1,1}}\frac{\d^{D}X}{(2\pi)^D}\,\exp\!\left[\im\,\sum_{j=1}^{n}\alpha_j\,\omega_j\,q_j\cdot X\right] 
    \\
    &= \frac{1}{(2\pi)^D}\int\limits_{\R^{d+1,1}} \d^DX\, \prod_{i=1}^n \frac{\mathrm{i}\,\alpha_i}{2\pi\,(u_i+q_i \cdot X)} \, , \label{eq:S_ndX}
\end{split}
\end{equation}
with the second line following by \eqref{eq:Fourierpw}. Each factor in the integrand is a $\Delta=1$ Carrollian conformal primary wavefunction at the same bulk point $X$, or equivalently, a Minkowski space bulk-to-boundary propagator from $X\in\R^{d+1,1}$ in the bulk to $(u_i,\mathbf{x}_i)\in\scri^{\alpha_i}$ on the boundary. This establishes that $S_{n}$ is the $n$-point contact diagram for Carrollian scalars.

While \eqref{eq:S_ndX} is still in the form of an integral, it is possible to obtain exact expressions for these Carrollian contact diagrams. There are two distinct cases to consider: $n>D$ and $n\leq D$, both of which can be evaluated along similar lines to their celestial counterparts~\cite{Casali:2020uvr}. 

\medskip

\paragraph{Case 1 $n>D$:} Rather than starting from the `obviously contact' representation \eqref{eq:S_ndX} of $S_n$, consider instead \eqref{eq:Sn}, where $S_n$ is written as the Fourier transform of the momentum conserving delta function. When $n>D$, these momentum conserving delta functions can be used to solve for $\omega_1,\ldots,\omega_D$ in terms of $\omega_{D+1},\ldots,\omega_n$. 

By decomposing the argument of the $D$-dimensional delta function using a basis of vectors
\be\label{basisdecomp}
\epsilon_{\mu\nu_2\cdots\nu_D}\,q_2^{\nu_2}\cdots q_D^{\nu_D}\,,\ldots,\,\epsilon_{\nu_1\cdots\nu_{D-1}\mu}\,q_1^{\nu_1}\cdots q_{D-1}^{\nu_{D-1}}\,,
\ee
one finds that 
\be\label{deltadecomp1}
\delta^{D}\!\left(\sum_{i=1}^{n}\alpha_i\,\omega_i\,q_i\right)=\frac{1}{(1\,2\cdots D)}\,\prod_{j=1}^{D}\delta(\omega_j-\omega_j^*)\,,
\ee
for
\be\label{deltadecomp2}
    \omega_j^*:=\sum_{r=D+1}^n V_{jr}\, \omega_r , \qquad V_{jr} := -\alpha_j\,\alpha_r \frac{(1\, 2\dots j-1\, r\, j+1 \dots D)}{(1\,2\,\cdots D)} \, ,
\end{equation}
with the notation for the determinant of $q_i$,
\begin{equation}
    (i_1 \, i_2\, \cdots i_D) := \epsilon_{\mu_1\mu_2\dots\mu_D}\,q_{i_1}^{\mu_1}\,q_{i_2}^{\mu_2}\cdots q_{i_D}^{\mu_D} \, .
\end{equation}
Demanding that $\omega^*_j>0$ for all $j=1,\ldots,D$ when $\omega_r>0$ for all $r=D+1,\ldots,n$ restricts all coefficients $V_{jr}$ in \eqref{deltadecomp2} to be positive (see e.g., Section 7 of~\cite{Mason:2023mti} for a discussion of the allowed kinematic region).

Using \eqref{deltadecomp1}, the first $D$ Fourier integrals in $S_n$ can be performed against the delta functions to leave
\begin{equation}
    S_n=\frac{1}{(2\pi)^n\,|(1\,2\cdots D)|} \prod_{j,r}\Theta(V_{jr})\,\left(\prod_{r=D+1}^{n} \int_{0}^{\infty} \d\omega_r\, \e^{\mathrm{i}\,\omega_r\, L_r}\right) \,,
\end{equation}
where 
\begin{equation}
    L_r := \alpha_r\, u_r +\sum_{j=1}^D \alpha_j V_{jr}\, u_j \, .
\end{equation}
Evaluating the remaining integrals directly (with an implicit regulator, corresponding to an $\im\varepsilon$-prescription, to damp the contributions at large frequencies) gives
\begin{equation}
     S_n=\frac{1}{(2\pi)^n\,|(1\,2\cdots D)|} \prod_{j,r}\Theta(V_{jr})\, \prod_{r=D+1}^n\frac{\mathrm{i}}{L_r} \, .
\end{equation}
This simple form of the Carrollian contact diagram has been found before (e.g., Example 2 in Section 7 of~\cite{Mason:2023mti}). There are other kinematically allowed regions of integration (e.g., Example 1 in Section 7 of~\cite{Mason:2023mti}), but we do not consider them, preferring to work with strictly positive frequencies.

\medskip

\paragraph{Case 2 $n\leq D$:} In this case, the momentum conserving delta functions in \eqref{eq:Sn} can be used to solve for $\omega_1,\ldots,\omega_{n-1}$ in terms of $\omega_n$. To that end, partition spacetime indices $\mu$ into two sets: $\mu=(a,b)$ where $a=0,1,\ldots,D-n$, $b=D-n+1,\ldots,D-1$. 

Consider the final $n-1$ momentum conserving constraints:
\begin{equation}
    \sum_{j=1}^{n} \alpha_j\,\omega_j\, q_j^b = -\alpha_n\, \omega_n\, q_n^b\,.
\end{equation}
These can be solved for each $\omega_1,\ldots,\omega_{n-1}$ by contracting with the basis vectors
\be\label{deltadecomp3}
\epsilon_{aa_2\cdots a_{n-1}}\,q_2^{a_2}\cdots q_{n-1}^{a_{n-1}}\,,\ldots,\,\epsilon_{a_1\cdots a_{n-2}a}\,q_{1}^{a_1}\cdots q_{n-2}^{a_{n-2}}\,,
\ee
to give
\begin{equation}
    \omega_j= V_{jn}\,\omega_n \, , \qquad V_{jn} := -\alpha_j\,\alpha_n \frac{(12\cdots j-1\, n \, j+1 \cdots n-1)}{(12\cdots n-1)} \, ,
\end{equation}
where the determinant is now defined as
\begin{equation}
    (i_1 i_2 \cdots i_{n-1}) := \epsilon_{a_1 a_2 \cdots a_{n-1}}\,q_{i_1}^{a_1}\, q_{i_2}^{a_2}\cdots q_{i_{n-1}}^{a_{n-1}} \, .
\end{equation}
in terms of the $(n-1)$-dimensional Levi-Civita symbol. Once again, the requirement that all frequencies be positive carves out the region of kinematic space where all of the coefficients $V_{jn}$ are positive.

The expression for $S_n$ now becomes
\begin{multline}\label{nDcont1}
      S_n = \delta^{D-n+1}\!\left(\sum_{j=1}^{n-1} \alpha_j\,V_{jn}\,q_j^{a} +\alpha_n\, q^{a}_n\right) \frac{1}{(2\pi)^n\,|(12\cdots n-1)|}\, \prod_{j=1}^{n-1} \Theta(V_{jn})\\
     \times\, \int_0^\infty \frac{\d\omega_n}{\omega_n^{D-n+1}}\, \exp\!\left[\mathrm{i}\,\omega_n\left(\alpha_n\, u_n+\sum_{j=1}^{n-1}\alpha_j\, V_{jn}\,u_j\right)\right] \, .
\end{multline}
The remaining $\omega_n$ integral is IR divergent (as $D-n+1\geq1$). However, one can regularize it by computing $\partial_u$-descendants; for instance,
\begin{multline}\label{nDcont2}
     \partial_{u_n}^{D-n+1} S_n =  \delta^{D-n+1}\!\left(\sum_{j=1}^{n-1} \alpha_j\,V_{jn}\,q_j^{a} +\alpha_n\, q^{a}_n\right) \frac{1}{(2\pi)^n\,|(12\cdots n-1)|}\, \prod_{j=1}^{n-1} \Theta(V_{jn}) \\
     \times\,\frac{\mathrm{i}(\mathrm{i}\,\alpha_n)^{D-n+1}}{\alpha_n\, u_n+\sum_{j=1}^{n-1} \alpha_j\, V_{jn}\,u_j} \, . 
\end{multline}
This sort of regularization by taking $u$-derivatives is typical of Carrollian amplitudes (cf., \cite{Donnay:2022wvx,Mason:2023mti}). Note that while splitting the tangent space into effective subspaces of dimension $D-n+1$ and $n-1$ apparently breaks Poincar\'e invariance, the delta functions in \eqref{nDcont1}, \eqref{nDcont2} ensure that the final answer is actually invariant.


\subsection{Ambitwistor strings in the Carrollian basis}

The CHY formulae can be obtained from worldsheet correlation functions of \emph{ambitwistor string theories}~\cite{Mason:2013sva,Casali:2015vta}, which govern holomorphic maps from a closed Riemann surface to the space of scale-less, complexified null geodesics of spacetime~\cite{Isenberg:1978kk,Witten:1978xx,LeBrun:1983,Witten:1985nt}. The matter content of the worldsheet theory dictates which CHY integrands are produced, and thus which amplitudes are obtained. Furthermore, the scattering equations emerge from the worldsheet path integral, and have the geometric interpretation of ensuring that ambitwistor space is the target space~\cite{Mason:2013sva,Adamo:2013tsa}.

These results all followed from representing the vertex operators in the ambitwistor string BRST cohomology in a momentum eigenstate basis. If instead one uses a Carrollian primary basis, then we expect to obtain the Carrollian CHY formulae \eqref{CarCHY2}, with the operator-valued Carrollian scattering equations. Here, we verify that this is indeed the case, providing a consistency check on our formulae for all-multiplicity Carrollian scattering amplitudes.

\medskip

After gauge-fixing, all ambitwistor strings have an action of the form:
\be\label{ATS1}
S=\frac{1}{2\pi}\int_{\Sigma}P_{\mu}\,\dbar X^{\mu}+b\,\dbar c+\tilde{b}\,\dbar\tilde{c}+S_{\mathrm{matter},\,\mathrm{ghosts}}\,,
\ee
and associated BRST charge
\be\label{ATS-BRST}
Q=\frac{1}{2\pi\im}\oint c\,T+b\,c\,\partial c+\frac{\tilde{c}}{2}\,P^2+J^{\mathrm{matter}}\,.
\ee
Here, $X^{\mu}$ is a map from the closed Riemann surface $\Sigma$ to $\C^{D}$ (the complexification of $\R^{d+1,1}$), while the conjugate worldsheet field $P_\mu$ carries conformal weight $(1,0)$ (i.e., it is a section of the canonical bundle $K_{\Sigma}$). Conformal gauge has been chosen so that $\dbar=\d\bar{z}\,\partial_{\bar{z}}$ is the anti-holomorphic Dolbeault operator on $\Sigma$. The fermionic ghost systems $(c,b)$, $(\tilde{c},\tilde{b})$ have identical conformal weights --  $(-1,0)$ for $c,\,\tilde{c}$ and $(2,0)$ for $b,\,\tilde{b}$ -- and correspond to holomorphic conformal transformations and the gauge freedom conjugate to the ambitwistor constraint $P^2=0$, respectively. $S_{\mathrm{matter},\,\mathrm{ghosts}}$ denotes any additional worldsheet matter and associated ghosts arising from gauge-fixing.

In the BRST charge \eqref{ATS-BRST}, normal-ordering is implicitly assumed; $T$ denotes the full holomorphic stress tensor (including contributions from all matter and ghost fields), while $J^{\mathrm{matter}}$ denotes further contributions to the charge coming from specific choices of matter. For example, the choices of matter which eventually lead to the CHY formulae for biadjoint scalars, Yang-Mills theory and type II supergravity are, respectively:
\be\label{ATS-Type0}
S_{\mathrm{BAS}}=S_{\mathfrak{g}}+S_{\mathfrak{\tilde{g}}}\,, \qquad J^{\mathrm{BAS}}=0\,,
\ee
where $S_{\mathfrak{g},\tilde{\mathfrak{g}}}$ are actions for worldsheet current algebras for semi-simple Lie algebras $\mathfrak{g}$, $\tilde{\mathfrak{g}}$;
\be\label{ATS-Het}
S_{\mathrm{YM}}=S_{\mathfrak{g}}+\frac{1}{2}\,\Psi_{\mu}\,\dbar\Psi^{\mu}+\beta\,\dbar\gamma\,, \quad J^{\mathrm{YM}}=\gamma\,\Psi\cdot P+\frac{\tilde{b}}{2}\,\gamma^2\,,
\ee
where $\Psi_{\mu}$ are fermionic fields of conformal weight $(\frac{1}{2},0)$ and $(\gamma,\beta)$ are bosonic ghosts of conformal weight $(-\frac{1}{2},0)$, $(\frac{3}{2},0)$; and
\be\label{ATS-TypeII}
\begin{split}
S_{\mathrm{II}}=\frac{1}{2}\left(\Psi_{\mu}\,\dbar\Psi^{\mu}+\tilde{\Psi}_{\mu}\,\dbar\tilde{\Psi}^{\mu}\right)+\beta\,\dbar\gamma+\tilde{\beta}\,\dbar\tilde{\gamma} \\
J^{\mathrm{II}}=\gamma\,\Psi\cdot P+\tilde{\gamma}\,\tilde{\Psi}\cdot P +\frac{\tilde{b}}{2}\left(\gamma^2+\tilde{\gamma}^2\right)\,,
\end{split}
\ee
where the tilded fields have the same statistics and conformal weights as their un-tilded counterparts.

While there are many potential anomalies obstructing $Q^2=0$, only the holomorphic conformal anomaly, controlled by the central charge, is generically non-zero. This anomaly is entirely controlled by the target space dimension, $D$, which can always be chosen to render the theory anomaly-free. For instance, the ambitwistor string with no additional matter content is anomaly free for $D=26$, while the theory corresponding to type II supergravity has critical dimension $D=10$~\cite{Mason:2013sva}. However, when attention is confined to genus zero (i.e., $\Sigma\cong\CP^1$), sensible worldsheet correlators can be computed for generic $D$, as the conformal anomaly does not affect the final correlator formulae on the sphere, and the CHY formulae \eqref{eq:CHY1} in arbitrary dimension are obtained. 

\medskip

To obtain Carrollian amplitudes from ambitwistor strong correlators, one should simply express vertex operators in the BRST cohomology in terms of Carrollian primaries. The derivation for biadjoint scalar theory is sufficient to demonstrate the general principles, so we focus on this simplest case for clarity. In the worldsheet theory defined by \eqref{ATS-Type0}, a generic ansatz for fixed vertex operators is
\be\label{fixedVO}
V=c\,\tilde{c}\,j\cdot\mathsf{T}\,\tilde{j}\cdot\tilde{\msf{T}}\,\phi(X)\,,
\ee
where $j^{\msf{a}}$, $\tilde{j}^{\tilde{\msf{a}}}$ are the conformal weight $(1,0)$ Kac-Moody currents for $\mathfrak{g}$, $\tilde{\mathfrak{g}}$, respectively, and $\phi$ is a yet-to-be-determined function of $X$. The non-singular worldsheet OPEs are
\be\label{OPEs}
P_{\mu}(z)\,X^{\nu}(w)\sim \frac{\delta^{\nu}_{\mu}}{z-w}\,, \qquad b(z)\,c(w)\sim\frac{1}{z-w}\,,
\ee
\be\label{KM-OPE}
    j^{\msf{a}}(z)\,j^{\msf{b}}(w)\sim \frac{\kappa\,\delta^{\msf{ab}}}{(z-w)^2}+\frac{f^{\msf{abc}}\,j^{\msf{c}}(w)}{z-w}\,,
\ee
and similar for the tilded fields, where $\kappa$ and $f^{\msf{abc}}$ are the level of the worldsheet current algebra for $\mathfrak{g}$ and the structure constants, respectively. The double pole in \eqref{KM-OPE}, controlled by the levels $\kappa,\tilde{\kappa}$, leads to multi-trace terms in correlation functions. As our interest will be restricted to the single-trace sector, it is convenient to formally set $\kappa=0=\tilde{\kappa}$ when computing OPEs between Kac-Moody currents; this has the practical effect of removing the multi-trace terms~\cite{Berkovits:2004jj,Azevedo:2017lkz,Adamo:2018hzd}\footnote{Alternatively, one could employ the remarkable recent construction of~\cite{Seet:2025mes} to remove multi-trace contributions while still working with finite level.}. 

With these OPEs, it follows that
\be\label{Qclose1}
Q\,V=\frac{c\,\tilde{c}\,\partial\tilde{c}}{2}\,j\cdot\msf{T}\,\tilde{j}\cdot\tilde{\msf{T}}\,\Box\phi(X)\,,
\ee
so that $V$ is in the BRST cohomology if $\phi$ is a solution to the massless wave equation in $D$-dimensional flat space. So, rather than representing the vertex operator with a plane wave momentum eigenstate, we are free to use the Carrollian conformal primary wavefunctions \eqref{eq:Fourierpw}, giving fixed vertex operators\footnote{It should be emphasized that these are \emph{not} the only fixed vertex operators in the theory, which contains many other states in its massless spectrum besides massless biadjoint scalars (cf., \cite{Berkovits:2018jvm,Figueroa-OFarrill:2024wgs}).} of the form
\be\label{fixedVO2}
V=c\,\tilde{c}\,j\cdot\mathsf{T}\,\tilde{j}\cdot\tilde{T}\,\phi^{\alpha}(X;u,q)\,,
\ee
where we recall that $\alpha=\pm1$ denotes outgoing/incoming solutions and $(u,q)$ labels a point on $\scri^{\alpha}$; in particular, $(u,q)$ are quantum numbers, \emph{not} worldsheet fields.

We can now compute a correlation function of $n$ such vertex operator insertions on a genus zero worldsheet. In the presence of vertex operator insertions, there are obstructions to achieving the gauge fixing \eqref{ATS1} at all of the vertex operator insertions; these obstructions correspond to the cohomology group $H^1(\Sigma,T_{\Sigma}(-z_1-\cdots-z_n))$, where sections of $T_{\Sigma}(-z_i)$ are sections of the holomorphic tangent bundle of $\Sigma$ with simple zeroes at $z_i$, the location of the $i^{\mathrm{th}}$ vertex operator insertion on $\Sigma$. At genus zero, there are $n-3$ such obstructions, which can be parametrized by a basis of the form
\be\label{cohbasis}
e_i=\frac{(z-z_i)}{\d z}\,\dbar\left(\frac{1}{z-z_i}\right)\,\frac{(z-z_1)\,(z-z_2)\,(z-z_3)}{z_{i1}\,z_{i2}\,z_{i3}}\,, \quad i=4,\ldots,n\,.
\ee
These objects are only non-trivial as distributions when integrated against quantities that have poles at $z=z_i$.

The worldsheet correlator is then defined by
\be\label{WSCorr1}
\left\la\left[\prod_{i=4}^{n}\int \frac{\d r_i}{(2\pi\im)^2}\,\exp\!\left(-\frac{1}{4\pi}\int_{\Sigma}r_i\,e_i\,P^2\right)\left(\int_{\Sigma}e_i\,b\right) \left(\int_{\Sigma}e_i\,\tilde{b}\right)\right] \prod_{j=1}^{n}V_j\right\ra\,,
\ee
with the integral over $\{r_i\}$ corresponding to summing over possible linear combinations of the basis elements \eqref{cohbasis}. The insertions of $\int e_i b$ and $\int e_i\tilde{b}$ ensure that the correct number of zero modes for the fermionic ghost fields $c,\tilde{c}$ will be included in the path integral to give a non-vanishing result. Now, using the OPEs \eqref{OPEs} and the definition of the basis elements \eqref{cohbasis}, it follows that
\be\label{WSCorr2}
\int_{\Sigma}e_i\,b\,c(z_j)=2\pi\im\,\delta_{ij}\,,
\ee
and similarly for the contractions between $\int e_i\tilde{b}$ and $\tilde{c}(z_j)$. This reduces the correlator to
\begin{multline}\label{WSCorr3}
\left\la c(z_1)c(z_2)c(z_3)\,\tilde{c}(z_1)\tilde{c}(z_2)\tilde{c}(z_3)\left[\prod_{i=4}^{n}\int \d r_i\,\exp\!\left(-\frac{1}{4\pi}\int_{\Sigma}r_i\,e_i\,P^2\right)\right]\right. \\ \times\,\left.\prod_{j=1}^{n}j(z_j)\cdot\msf{T}_j\,\tilde{j}(z_j)\cdot\tilde{\msf{T}}_j\,\phi_j\right\ra\,,
\end{multline}
where $\phi_i\equiv\phi^{\alpha_i}(X(z_i);u_i,q_i)$.

Since the remaining $c,\tilde{c}$ ghost fields have no further anti-ghost fields to contract with, their contribution to the correlator is fixed by restricting them to their zero-mode values. Both $c$ and $\tilde{c}$ have conformal weight $(-1,0)$ and thus 3 zero modes at genus zero; this is precisely the number required to ensure that their fermionic path integral contributing to \eqref{WSCorr3} is non-vanishing:
\be\label{WSCorr4}
\left\la c(z_1)c(z_2)c(z_3)\,\tilde{c}(z_1)\tilde{c}(z_2)\tilde{c}(z_3)\right\ra=\frac{z_{12}^2\,z_{23}^2\,z_{31}^2}{\d z^2_1\,\d z^2_2\,\d z^2_3}\,.
\ee
Furthermore, the single-trace contribution from the Kac-Moody currents is:
\be\label{WSCorr5}
\left\la \prod_{j=1}^{n}j\cdot\msf{T}\,\tilde{j}\cdot\tilde{\msf{T}}\right\ra=\mathrm{PT}_n\,\widetilde{\mathrm{PT}}_n\,\prod_{j=1}^{n}\d z_j^2\,,
\ee
with the factors $\mathrm{PT}_n$, $\widetilde{\mathrm{PT}}_n$ defined by \eqref{PTfact}.

This leaves
\be\label{WSCorr6}
\left\la z_{12}^2\,z_{23}^2\,z_{31}^2 \left[\prod_{i=4}^{n}\int \d z_i^2\,\d r_i\,\exp\!\left(-\frac{1}{4\pi}\int_{\Sigma}r_i\,e_i\,P^2\right)\right]\mathrm{PT}_n\,\widetilde{\mathrm{PT}}_n\, \prod_{j=1}^{n}\phi_j\right\ra\,,
\ee
with only the $XP$ system's contribution to the correlator remaining. Recalling the form of the Carrollian wavefunctions \eqref{eq:Fourierpw} and the OPEs \eqref{OPEs}, it follows that
\be\label{HamOPE1}
P^2(z)\,\prod_{j=1}^{n}\phi_j\sim 2\,\d z^2\,\sum_{j=1}^{n}\frac{\partial_\mu\phi_j}{z-z_j}\,\sum_{k\neq j}\frac{\partial^{\mu}\phi_k}{z-z_k}\,\prod_{i\neq j,k}\phi_{i}\,,
\ee
while
\be\label{HamOPE2}
\partial_\mu \phi_j=\frac{-\im\,\alpha_j\,q_{j\,\mu}}{2\pi\,(u_j+q_j\cdot X)^2}=q_{j\,\mu}\,\partial_{u_j}\phi_j\,.
\ee
Thus, we see that
\be\label{HamOPE3}
P^2(z)\,\prod_{j=1}^{n}\phi_j\sim2\,\d z^2\left(\sum_{\substack{k,l=1 \\ k\neq l}}^{n}\frac{\mathbf{K}_k\cdot\mathbf{K}_l}{(z-z_k)\,(z-z_l)}\right)\prod_{j=1}^{n}\phi_j\,,
\ee
where $\mathbf{K}_i^{\mu}$ is the differential operator \eqref{Kdef}.

Since there is no other $X$-dependence in the path integral, we can replace $P^2$ in the exponentials of \eqref{WSCorr6} with the differential operator in \eqref{HamOPE3}, and set $X^{\mu}(z)=X^{\mu}$, its constant zero mode which must be integrated over. This gives
\be\label{WSCorr7}
-\frac{1}{4\pi}\int_{\Sigma}r_i\,e_i\,P^2=\im\,r_i\,\mathbf{E}_i\,\d z_i\,,
\ee
in the argument of the exponentials, where $\mathbf{E}_i$ is the $i^{\mathrm{th}}$ Carrollian scattering `equation' (really a differential operator). At this point, the evaluation of the correlator is finally complete; after rescaling $r_i\to r_i\,\d z_i^{-1}$ one obtains
\be\label{WSCorr8}
\int\limits_{\R^{d+1,1}}\d^{D}X\,z_{12}^2\,z_{23}^2\,z_{31}^2\,\left(\prod_{i=4}^{n}\int \d z_i\,\d r_i\,\e^{\im\,r_i\,\mathbf{E}_i}\right)\mathrm{PT}_n\,\widetilde{\mathrm{PT}}_n\,\prod_{j=1}^{n}\phi_{j}\,.
\ee
Using \eqref{SL2quot} and \eqref{eq:S_ndX}, this can be written in a more covariant fashion as
\be\label{WSCorr9}
 \int\frac{\d^{n}z}{[\mathrm{vol}\,\SL(2,\C)]^2}\,\left(\prod_{i=1}^{n}\int_{-\infty}^{\infty}\frac{\d r_i}{2\pi}\,\e^{\im\,r_i\,\mathbf{E}_i}\right)\,\mathrm{PT}_n\,\widetilde{\mathrm{PT}}_n\,S_{n}\,,
\ee
up to an irrelevant overall numerical factor. This is recognized as the Carrollian CHY formula for the tree-level S-matrix of biadjoint scalar theory, as desired. It is straightforward to generalise this argument to ambitwistor strings with other types of matter, where one obtains more general Carrollian CHY formulae of the form \eqref{CarCHY2}.

\section{Scattering equations in AdS} \label{sec:3}

One key application of Carrollian amplitudes is that they are naturally adapted to flat limits of boundary correlation functions in anti-de Sitter (AdS) space. While we now have expressions for all-multiplicity Carrollian amplitudes in any number of spacetime dimensions for a wide variety of massless QFTs, it might seem a difficult task to connect these with flat limits of AdS correlators. Indeed, to do so would require analogous, all-multiplicity formulae for tree-level boundary correlators in AdS of general dimension.

One could hope to use ambitwistor strings as a mechanism to generate such formulae, by changing the target space of the worldsheet theory from flat space to AdS. In fact, it \emph{is} possible to couple ambitwistor strings to curved background fields: unlike in ordinary string theories, the resulting anomaly cancellation conditions can be determined \emph{exactly}, and are given by the expected non-linear field equations~\cite{Adamo:2014wea,Adamo:2018hzd}. Furthermore, the BRST cohomology leads to vertex operators which are determined by massless free fields coupled to the non-linear background~\cite{Adamo:2018ege}. 

However, in the gravitational case, the equations of motion are those of NS-NS supergravity, for which AdS is not a solution. As usual, it is possible to support the negative scalar curvature of AdS by switching on background Ramond-Ramond fluxes, but this is impossible within the RNS-like formalism of the ambitwistor string described here. While a pure spinor formalism for the ambitwistor string is known~\cite{Berkovits:2013xba,Gomez:2013wza,Adamo:2015hoa}, it is only known how to \emph{classically} couple the worldsheet theory to background supergravity fields (cf., \cite{Chandia:2015xfa}). Consequently, it seems like the task of computing ambitwistor string correlators to obtain AdS boundary correlation functions is dead in the water.

Despite these apparent obstructions, it is a remarkable fact that simply ignoring the various anomalies and computing na\"ive genus zero correlators in the ambitwistor string for biadjoint scalars in AdS results in \emph{correct}, all-multiplicity answers in any number of dimensions~\cite{Eberhardt:2020ewh,Roehrig:2020kck}! While this na\"ive approach fails to yield correct results for other (non-scalar) theories, it still provides a wealth of data which can be used to study flat-space limits. 

\medskip

Recall that $D$-dimensional AdS can be described as a quadratic hypersurface in $D+1$ dimensional embedding space: if $\cX^{A}$ are coordinates on $\R^{D-1,2}$ with metric
\be\label{embedding1}
\d s^{2}_{\R^{D-1,2}}=\sum_{A=1}^{D}(\d\cX^{A})^2-(\d\cX^0)^2-(\d\cX^{-1})^{2}\,,
\ee
then (Lorentzian) AdS$_{D}$ is defined by the quadric
\be\label{embedding2}
\cX\cdot\cX=-\ell^2\,,
\ee
where $\ell$ is the AdS radius. Points on the AdS boundary correspond to points on the projective null cone:
\be\label{embedding3}
\partial\mathrm{AdS}=\left\{K^{A}\in\R^{D-1,2}\,|\,K^2=0\right\}/\Upsilon\,,
\ee
where the quotient is by the Euler vector $K\cdot\frac{\partial}{\partial K}$ which generates projective rescalings $K^{A}\to a\,K^{A}$ for any non-vanishing $a\in\R^*$.

Among the many advantages of working in embedding space are the simplicity of bulk-to-boundary propagators and conformal generators when written in the embedding coordinates. In particular, the bulk-to-boundary propagator for a boundary operator of scaling dimension $\Delta$ is
\be\label{b2b}
\Phi_{\Delta}(\cX;K)=\ell^{\Delta-\frac{D-2}{2}}\,\frac{\tilde{C}_2(\Delta)}{\left(-2\,K\cdot\cX\right)^{\Delta}}\,,
\ee
with the normalisation constant
\be\label{C2norm}
\tilde{C}_{2}(\Delta)=\frac{-\im\,(-1)^{D}}{2}\,\frac{\Gamma(\Delta)}{\pi^{\frac{D-1}{2}}\,\Gamma(\Delta-\frac{D-3}{2})}\,\times\,\left\{\begin{array}{rl}
                                                     1 & \mbox{ if } \Delta\neq\frac{D-2}{2}  \\
                                                     \frac{1}{2} & \mbox{ if } \Delta=\frac{D-2}{2}
                                                     \end{array}\right.\,,
\ee
while the generators of the boundary conformal algebra are 
\be\label{ConfGens}
\cD^{AB}:=K^{A}\,\frac{\partial}{\partial K_B}-K^{B}\,\frac{\partial}{\partial K_A}\,.
\ee
Surprisingly, these are the only ingredients required to obtain a CHY-like formula for all tree-level boundary correlation functions of biadjoint scalar theory in AdS$_D$, with arbitrary mass.

Specifically, consider the classical theory theory:
\begin{equation}
    S[\phi]=\int_{\mathrm{AdS}_D} D\phi_{\msf{a}\tilde{\msf{a}}}\wedge \star D\phi^{\msf{a}\tilde{\msf{a}}}-m^2\,\star\phi_{\msf{a}\tilde{\msf{a}}}\phi^{\msf{a}\tilde{\msf{a}}}+\frac{f^{\msf{abc}}\,f^{\tilde{\msf{a}}\tilde{\msf{b}}\tilde{\msf{c}}}}{3}\,\star \phi^{\msf{a}\tilde{\msf{a}}}\,\phi^{\msf{b}\tilde{\msf{b}}}\,\phi^{\msf{c}\tilde{\msf{c}}}\,, \label{eq:Sphi}
\end{equation}
where $\phi^{\msf{a}\tilde{\msf{a}}}$ is the scalar field living in the adjoint representations of two groups $G$, $\tilde{G}$, with $f^{\msf{abc}}$, $f^{\tilde{\msf{a}}\tilde{\msf{b}}\tilde{\msf{c}}}$ the structure constants of the corresponding Lie algebras $\mathfrak{g}$, $\tilde{\mathfrak{g}}$, and $\star$ is the AdS$_D$ Hodge star. This is simply the usual biadjoint scalar theory, but now including a mass, which can be related to an AdS boundary scaling dimension by
\be\label{AdSmass}
m^2=\frac{\Delta}{\ell^2}\,(\Delta-D+1)\,,
\ee
via the usual analysis of AdS wave equations (cf., \cite{Gubser:1998bc,Witten:1998qj}).

\medskip

By formulating the ambitwistor string with two worldsheet current algebras \eqref{ATS-Type0} with AdS$_D$ target space, vertex operators for these massive biadjoint scalars can be given by
\be\label{AdSVO}
V=c\,\tilde{c}\,j\cdot\msf{T}\,\tilde{j}\cdot\tilde{\msf{T}}\,\Phi_{\Delta}(\cX;K)\,,
\ee
where $\cX$ is now a worldsheet field while $K$ is a fixed quantum number. While the worldsheet model is wildly anomalous, na\"ive computation of genus zero correlators (along with a bit of intuition for the $m\neq0$, $\Delta\neq D-1$ case~\cite{Gomez:2021qfd,Gomez:2021ujt,Armstrong:2022csc}) leads to the formula~\cite{Eberhardt:2020ewh,Roehrig:2020kck}:
\begin{equation}
    \mathcal{C}_{n}(\Delta;\ell)=\int\frac{\d^{n}z}{[\mathrm{vol}\,\SL(2,\C)]^2}\,\mathrm{PT}_n\,\widetilde{\mathrm{PT}}_n\left(\prod_{i=1}^{n}\int_{-\infty}^{\infty}\frac{\d r_i}{2\pi} \e^{\im\,r_i\,\mathcal{E}_i}\right)\mathcal{S}_n\, , \label{eq:AdSCHY}
\end{equation}
where the differential operators
\be\label{AdSSE}
\mathcal{E}_i:=2\,\sum_{j\neq i}\frac{\ell^{-2}\,\mathcal{D}_i\cdot\mathcal{D}_j-m^{2}_{ij}}{z_{ij}}\,, \qquad m^2_{ij}=\left\{\begin{array}{rl}
                                m^2=\ell^{-2}\Delta(\Delta-D+1) & \mbox { if } j=i\pm1 \\
                                0 & \mbox{ otherwise }
                                \end{array}\right.\,,
\ee
are the AdS scattering equations, and
\be\label{AdScontact}
\mathcal{S}_n:=\int_{AdS_{D}}\d^{D+1}\cX\,\delta(\cX^2+\ell^2)\,\prod_{i=1}^{n}\Phi_{\Delta}(\cX;K_i)\,,
\ee
is the $n$-point scalar contact Witten diagram~\cite{Freedman:1998tz}. Division by the two factors of the volume of SL$(2,\C)$ in \eqref{eq:AdSCHY} is meant in the same sense as the flat space CHY formulae \eqref{SL2quot}.

Despite the somewhat dubious worldsheet origin of \eqref{eq:AdSCHY}, it can nevertheless be shown that $\mathcal{C}_{n}(\Delta;\ell)$ is the correct $n$-point tree-level boundary correlator of massive biadjoint scalars in AdS$_{D}$. The proof boils down to showing that the action of the scattering equations on the contact diagram generates a Berends-Giele-like recursion for the sum of cubic Witten diagrams which build the boundary correlators~\cite{Eberhardt:2020ewh}. Since its initial formulation for massless biadjoint scalars, the AdS scattering equations and associated CHY formulae have also been formulated for momentum space correlators in de Sitter space, as well as for more generic scalar effective field theories~\cite{Gomez:2021qfd,Gomez:2021ujt,Armstrong:2022csc}.


\section{From AdS correlators to Carrollian amplitudes} \label{sec:4}

The all-multiplicity expressions for biadjoint scalar boundary correlators in AdS \eqref{eq:AdSCHY} and Carrollian amplitudes \eqref{WSCorr9} in flat space share some heuristic features, including the fact that both the AdS and Carrollian scattering `equations' are not actually algebraic equations, but rather differential operators. Furthermore, in both instances these scattering equations act on a universal factor of the $n$-point scalar contact diagram, \eqref{AdScontact} in AdS and \eqref{eq:S_ndX} in flat space. Yet in the details, the formulae are very different --  not least as the underlying spacetimes have distinct symmetries which are encoded in the correlators or amplitudes.

In this section, we show that, when expressed in an appropriate set of coordinates, the flat space limit of the CHY formula \eqref{eq:AdSCHY} for AdS boundary correlators gives the CHY formula \eqref{WSCorr9} for Carrollian amplitudes. Different choices of mass for the AdS theory simply lead to Carrollian descendants (i.e., derivatives with respect to Bondi time) of the flat space amplitudes. After reviewing the appropriate Bondi coordinate system in AdS, we carefully implement the flat space limit to recover the Carrollian amplitudes. The arguments hold in any number of spacetime dimensions and with an arbitrary number of external states. 


\subsection{Bondi coordinates for AdS}

Any particular AdS coordinate system is given by choosing a `section' of the embedding space coordinates; this is equivalent to a choice of affine coordinate patch for a projective space. Just as the Bondi coordinates \eqref{eq:gflatBondi} are natural for describing Carrollian amplitudes in flat space, there is Bondi coordinate system in AdS which is naturally adapted to describing the flat limit of boundary correlators~\cite{Barnich:2012aw,Poole:2018koa,Kulkarni:2025qcx}\footnote{These AdS-Bondi coordinates are also closely tied to alternative boundary conditions for asymptotically AdS spacetimes which allow for fluctuations of the spatial boundary metric and BMS-like asymptotic symmetry algebras~\cite{Compere:2019bua,Compere:2020lrt,Geiller:2022vto,Hartong:2025jpp,Fiorucci:2025twa}.}. The AdS-Bondi coordinates $(u,r,\mathbf{x})$ are given by the embedding space section
\begin{equation}\label{Bondisection}
    \begin{split}
    \cX^A&=\left(\cX^{-1},\,\cX^{0},\,\cX^{1},\ldots,\,\cX^{D}\right) \\
     &=r\left(\frac{1}{2}+\frac{u}{r}-\frac{u^2}{2\,\ell^2}+\frac{|\mathbf{x}|^2}{2},\,\frac{u}{\ell}-\frac{\ell}{r},\,\frac{1}{2}-\frac{u}{r}+\frac{u^2}{2\,\ell^2}-\frac{|\mathbf{x}|^2}{2},\, \mathbf{x} \right) \, ,
    \end{split}
\end{equation}
which is easily confirmed to obey $\cX\cdot\cX=-\ell^2$ in the embedding space metric. The AdS metric in these coordinates is
\begin{equation}\label{AdSBondi}
    \d s^2_{\mathrm{AdS}_D}= -\frac{r^2}{\ell^2}\,\d u^2 -2\,\d u\,\d r + r^2\, |\d\mathbf{x}|^2 \, ,
\end{equation}
which reduces to the flat metric \eqref{eq:gflatBondi} in the limit\footnote{Of course, $\ell$ is a dimensionful quantity, and the flat limit can be equivalently phrased in terms of the dimensionless quantity $r/\ell\to0$.} $\ell\rightarrow\infty$. 

As in flat space, the AdS-Bondi metric is conformally compactified with the conformal factor $R:=r^{-1}$, and the conformal boundary is located at $r\to\infty$, or $R=0^{\pm}$:
\be\label{AdSboundmet}
\d s^2_{\partial\mathrm{AdS}_D}=-\frac{\d u^2}{\ell^2}+|\d\mathbf{x}|^2\,.
\ee
As expected, the AdS$_D$ boundary has the Minkowski metric of $\R^{D-2,1}$; in the flat space limit $\ell\to\infty$, this reduces to the degenerate metric \eqref{eq:gninf} on $\scri$. In the embedding space, the relevant sections corresponding to AdS boundary points are defined by 
\begin{equation}
    K^A = \lim_{r\rightarrow\infty} \frac{\cX^A}{r} = \left(\frac{1}{2}-\frac{u^2}{2\,\ell^2}+\frac{|\mathbf{x}|^2}{2},\, \frac{u}{\ell},\,\frac{1}{2}+\frac{u^2}{2\,\ell^2}-\frac{|\mathbf{x}|^2}{2},\, \mathbf{x} \right) \, , \label{eq:defPI}
\end{equation}
which is easily seen to obey $K\cdot K=0$. 


\subsection{Flat limit}

Our goal is now to implement the flat limit at the level of the AdS scattering equations formula \eqref{eq:AdSCHY}. We proceed by considering each ingredient in this formula in turn. Firstly, note that the objects $\mathrm{PT}_n$, $\widetilde{\mathrm{PT}}_n$ are defined purely in terms of colour generators and marked points on the underlying Riemann sphere; as such, they are completely independent of the AdS scale $\ell$, and pass directly to flat space without modification. 

Let us then begin with the AdS$_D$ contact diagram \eqref{AdScontact}. First, note that\footnote{Throughout this section, our Bachmann-Landau `big $O$' notation is defined with respect to $\ell^{-1}$: $f(\ell)=O(\ell^{k})$ for some $k\in\Q$ if $f(\ell)\,\ell^{k}$ is bounded as $\ell\to\infty$.}
\be\label{B2b1}
 \begin{split}
-2\,K_i\cdot\cX&=-2\,(u_i-u)-\frac{r}{\ell^2}\,(u_i-u)^2+r\,|\mathbf{x}_i-\mathbf{x}|^2 \\
  & = -2\left(u_i+q_i\cdot X\right) + O(\ell^{-2})\,,
 \end{split}
\ee
which means that in the flat space limit
\be\label{B2b2}
\Phi_{\Delta}(\cX;K_i)=-\ell^{\Delta+\frac{2-D}{2}}\,\tilde{C}_{2}(\Delta)\,2\,\pi\,\im\,(-2)^{-\Delta}\,\alpha_i\,\left(\frac{\im\,\alpha_i}{2\pi\,(u_i+q_i\cdot X)^{\Delta}}\right)+O(\ell^{\Delta-\frac{D}{2}})\,.
\ee
The object in parenthesis is very nearly the Carrollian wavefunction \eqref{eq:Fourierpw}, but for the fact that the argument in the denominator is raised to the power $\Delta$. 

However, this can easily be expressed as
\be\label{Carrdesc1}
(u_i+q_i\cdot X)^{-\Delta}=\frac{(-1)^{\Delta}}{\Gamma(\Delta)}\,\frac{\partial^{\Delta-1}}{\partial u_i^{\Delta-1}}\left(\frac{1}{u_i+q_i\cdot X}\right)\,,
\ee
which is just a level $\Delta-1$ Carrollian descendant of the Carrollian primary wavefunction. In other words, \eqref{B2b2} becomes
\be\label{B2b3}
\Phi_{\Delta}(\cX;K_i)=\ell^{\Delta+\frac{2-D}{2}}\,\beta(\Delta)\,\alpha_i\,\partial^{\Delta-1}_{u_i}\,\phi^{\alpha_i}(X;u_i,q_i)+O(\ell^{\Delta-\frac{D}{2}})\,,
\ee
where 
\be\label{betadefn}
\beta(\Delta):=\frac{(-1)^{\Delta-1}\,2\pi\im\,\tilde{C}_{2}(\Delta)\,(-2)^{-\Delta}}{\Gamma(\Delta)}\,,
\ee
is an overall numerical constant inherited from the normalisation of the mass $m$ bulk-to-boundary propagator in AdS$_D$ and the massless Carrollian primary wavefunction in flat space.

Next, consider the AdS scattering equations \eqref{AdSSE}. The differential operators appearing in each term of $\mathcal{E}_{i}$ take the form
\be\label{AdSSE1}
\mathcal{D}_i\cdot\mathcal{D}_j=2\,K_i\cdot K_{j}\,\frac{\partial}{\partial K_i}\cdot\frac{\partial}{\partial K_{j}}-2\left(K_i\cdot\frac{\partial}{\partial K_j}\right)\left(K_j\cdot\frac{\partial}{\partial K_i}\right)\,,
\ee
in terms of derivatives on the boundary insertions of the AdS correlator. A short computation in the embedding space formalism shows that in AdS-Bondi coordinates
\be\label{AdSSE2}
K_{i}\cdot K_j=\frac{(u_i-u_j)^2}{2\,\ell^2}-\frac{(\mathbf{x}_i-\mathbf{x}_j)^2}{2}=q_i\cdot q_j+O(\ell^{-2})\,,
\ee
where $q_i\cdot q_j$ is understood to be the $D$-dimensional (rather than embedding space) Minkowski inner product. Similarly, by using the chain rule
\be\label{AdSSE3}
\frac{\partial}{\partial K_i^{A}}=\frac{\partial u_i}{\partial K_i^A}\,\frac{\partial}{\partial u_i}+\frac{\partial\mathbf{x}_i}{\partial K_i^A}\cdot\frac{\partial}{\partial\mathbf{x}_i}\,,
\ee
with \eqref{eq:defPI} it follows that
\be\label{AdSSE4}
\frac{\partial}{\partial K_i^{A}}=\delta_{A}^{0}\,\ell\,\frac{\partial}{\partial u_i}+O(\ell^0)\,.
\ee
This means that in the large $\ell$-limit
\be\label{AdSSE5}
\frac{\partial}{\partial K_i}\cdot\frac{\partial}{\partial K_{j}}=-\ell^2\,\partial_{u_i}\,\partial_{u_j}+O(\ell)\,, \qquad K_i\cdot\frac{\partial}{\partial K_j}=u_i\,\partial_{u_j}+O(\ell^{-1})\,,
\ee
and consequently that 
\be\label{AdSSE6}
\mathcal{D}_i\cdot\mathcal{D}_j=-\ell^2\,q_i\cdot q_j\,\partial_{u_i}\,\partial_{u_j}+O(\ell)=\ell^2\,\mathbf{K}_i\cdot\mathbf{K}_j+O(\ell)\,,
\ee
where $\mathbf{K}_i^{\mu}$ is the flat space differential operator \eqref{Kdef} for the $i^{\mathrm{th}}$ external state.

Recalling that the mass term in the AdS scattering equations \eqref{AdSmass} is order $\ell^{-2}$, it follows from \eqref{AdSSE6} that
\be\label{SEflatlim}
\mathcal{E}_i=\mathbf{E}_i+O(\ell^{-1})\,,
\ee
where $\{\mathbf{E}_i\}$ are precisely the Carrollian scattering equations \eqref{CarrSE} in flat space. We can now put this together with \eqref{B2b3}, and the fact that in the $\ell\to\infty$ limit the AdS-Bondi metric passes smoothly to the Minkowski metric in planar Bondi coordinates \eqref{eq:gflatBondi}. The result is
\be\label{FlatLimit0}
\boxed{\lim_{\ell\to\infty}\frac{\ell^{n\left(\frac{D-2}{2}-\Delta\right)}}{\beta(\Delta)^n}\,\mathcal{C}_{n}(\Delta;\ell)=\prod_{i=1}^{n}\alpha_i\,\partial_{u_i}^{\Delta-1}\,C_n\,,}
\ee
where $C_n$ is the $n$-point, tree-level Carrollian amplitude of biadjoint scalar theory in flat spacetime. While we used the scattering equations to deduce this, the final statement \eqref{FlatLimit0} is, of course, independent of the representation used for the boundary correlators and Carrollian amplitudes appearing on each side. In the end, \eqref{FlatLimit0} establishes an exact correspondence between (an appropriate definition of) the flat limit of tree-level boundary correlators of biadjoint scalar theory with arbitrary mass in AdS and tree-level Carrollian amplitudes of massless biadjoint scalar theory in Minkowski spacetime.

Thanks to the all-multiplicity, dimension-agnostic nature of formulae based on the scattering equations, this proof holds for \emph{any} number of external points and spacetime dimensions. Also, note that the mass of the scalar fields in AdS, controlled by $\Delta$ is imprinted on the resulting flat space scattering amplitude by the number of derivatives with respect to Bondi time, or Carrollian descent level, of the external states. For massless AdS states, where $\Delta=D-1$, the resulting Carrollian amplitude is still descended by level $D-2$ with respect to each external state; this mismatch with other claims in the literature (cf., \cite{Kulkarni:2025qcx}) is simply due to our convention that the Carrollian conformal primary wavefunctions are defined by the Fourier transform \eqref{eq:Fourierpw}, rather than the modified Mellin transform \eqref{modMellin}.

As the theories on each side of this correspondence are biadjoint scalar, the correlators and amplitudes on both sides of the correspondence are colour-ordered, composed of sums over distinct colour orderings in $\mathfrak{g}$ and $\tilde{\mathfrak{g}}$. However, by taking sums of the terms in these colour-ordered partial amplitudes, one can obtain the correlators of amplitudes of a simple cubic scalar theory. Since the equality \eqref{FlatLimit0} holds term-by-term in the sum over colour-orderings, it follows that this flat limit also holds for (non-coloured) cubic scalar theories more generally. 


\section{Comments on spinning correlators in AdS$_3$} \label{sec:5}

The flat limit of spinning AdS boundary correlators has long been known to be more subtle than that for scalars (see~\cite{Berenstein:2025qhb} for a recent discussion). Given how cleanly the scattering equations and CHY formulae manifest the flat limit for scalar correlators, it is natural to ask if we can say anything about spinning correlators using the same tools. Indeed, we have the full CHY dictionary of Carrollian amplitudes in flat space, including for gluons and gravitons.

The problem comes from the AdS side of the correspondence: in order for the ambitwistor string to be anomaly-free -- and therefore have a chance of producing a sensible CHY formula in AdS -- the AdS background must solve some non-linear equations of motion. For instance, for the ambitwistor string describing gravity \eqref{ATS-TypeII}, background gravitational fields must solve the equations of motion of type II supergravity in 10 spacetime dimensions~\cite{Adamo:2014wea,Berkovits:2018jvm}. On its own, AdS$_{D}$ does not solve these equations, and instead one typically encounters AdS$_{D}\times M_{10-D}$, where $M_{10-D}$ is a compact spatial manifold carrying some flux which supports the negative scalar curvature of AdS$_{D}$. This flux is typically carried by Ramond-Ramond fields, for which the non-linear description in the RNS-like formulation of the ambitwistor string is not known, making the corresponding worldsheet calculations prohibitively difficult.

However, in the special case of AdS$_3$, progress \emph{is} possible, as one can consider AdS$_3\times S^3$ supported by NS-NS 3-form flux through the $S^3$. As this only involves the NS-NS fields, it can be treated exactly within the RNS formalism (cf., \cite{Maldacena:2000hw}). In~\cite{Roehrig:2020kck}, this fact was exploited to formulate the ambitwistor string for AdS$_3\times S^3$, considering vertex operators which carry only AdS$_3$ quantum numbers (rendering the $S^3$ essentially a spectator at tree-level). Considering correlation functions of gluon vertex operators in a theory with a single worldsheet current algebra \eqref{ATS-Het}, a formula was obtained which the authors of~\cite{Roehrig:2020kck} conjectured to describe Yang-Mills-Chern-Simons theory in AdS$_{3}$, with Chern-Simons level equal to one. This was based on the observation that the 2-point correlator has the characteristic double pole arising from such a bulk Chern-Simons theory, meaning that the bulk gluon excitations are not massless (as in pure Yang-Mills), but instead have scaling dimension $\Delta=2$.

Parameterizing the bulk gluons in terms of holomorphic and anti-holomorphic boundary degrees of freedom (which is natural for a bulk Chern-Simons theory), the resulting correlator for the purely holomorphic sector is given by:
\begin{equation}\label{gluonholcorr}
    \mathcal{C}^{\text{hol}}_n=\int \frac{\mathrm{d}^n z}{[\text{vol SL}(2,\mathbb{C})]^2}\frac{\mathrm{PT}_n}{z_{12}}\left(\prod_{i=3}^n \epsilon_i\cdot\mathcal{T}_i\right)\frac{\epsilon_1\cdot \epsilon_2}{z_{12}}\prod_{j=1}^n \left(\int_{-\infty}^\infty \frac{\mathrm{d}r_j}{2\pi}\, \e^{\im\, r_j\, \mathcal{E}_j}\right)\,\mathcal{S}_n\,.
\end{equation}
Here, the AdS scattering equations $\mathcal{E}_i$ are the same as before \eqref{AdSSE} -- with $m=0$ -- and $\mathcal{S}_n$ is the contact diagram \eqref{AdScontact} with $\Delta=2$. The new ingredients relative to the biadjoint scalar case involve the gluon polarization vectors.

As the ambitwistor model is most easily formulated by treating AdS$_3$ as (a real slice of) the group manifold SL$(2,\C)$, the remaining ingredients in \eqref{gluonholcorr} are expressed in this language in the first instance. The natural coordinates on (complexified) AdS$_3$ in this case are $(\phi,\gamma,\tilde{\gamma})$ with metric
\begin{equation}
     \mathrm{d}s^2=\ell^2\left(\mathrm{d}\phi^2+\e^{2\phi}\,\mathrm{d}\gamma \mathrm{d}\tilde{\gamma}\right)\,,
\end{equation}
for which an element $g\in\SL(2,\C)$ is given by
\be\label{SL2element}
g(\phi,\gamma,\tilde{\gamma})=\e^{\phi}\left(\begin{array}{cc}
                        \gamma\,\tilde{\gamma}+\e^{-2\phi} & \tilde{\gamma}\\
                        \gamma & 1
                        \end{array}\right)\,.
\ee
The AdS$_3$ boundary corresponds to $\phi\to\infty$, with $(\gamma,\tilde{\gamma})$ acting as complex stereographic coordinates on the two-dimensional boundary (after the appropriate conformal rescaling).

In these variables, the holomorphic gluon polarization is chosen to be
\begin{equation}\label{AdSPol}
    \epsilon_a=\langle\lam|\mathrm{t}_a|\lam\rangle\,, \qquad  \langle \lam|:=(1,-\gamma)\,,
\end{equation}
where the $\mathfrak{sl}(2,\C)$ generators are:
\begin{equation}
    \mathrm{t}_0=\frac{1}{2}\begin{pmatrix} 1 & 0\\
    0 & -1
    \end{pmatrix},\ \mathrm{t}_+=\begin{pmatrix}
        0 & 1\\
        0 & 0
    \end{pmatrix},\ \mathrm{t}_-=\begin{pmatrix}
        0 & 0\\
        1 & 0
    \end{pmatrix}\,.
\end{equation}
With this polarization, the remaining ingredients in \eqref{gluonholcorr} are the differential operators $\epsilon_i\cdot\mathcal{T}_i$, which are given by
\be\label{edotT}
\epsilon_i\cdot \mathcal{T}_i=\sum_{j\neq i}\frac{\gamma_{ij}}{z_{ij}}\left(2-\gamma_{ij}\,\frac{\partial}{\partial \gamma_j}\right)\,,
\ee
where the first term arises from $\Delta=2$ for the external gluons.

\medskip

To investigate the flat limit of this formula, one simply translates the various ingredients to AdS-Bondi coordinates and looks for the leading behaviour as $\ell\to\infty$. One finds that, on the AdS boundary,
\be\label{SL2-Bondi}
\gamma=x-\frac{u}{2\,\ell}\,,
\ee
where $x$ is the coordinate which, in the flat limit, is along the celestial circle. 
In the flat limit, we find that the differential operator $\epsilon_i\cdot\mathcal{T}_i$ behaves as
\begin{equation}
  \epsilon_i\cdot \mathcal{T}_i=  -\ell\,\sum_{j\neq i}\frac{x_{ij}^2\,\partial_{u_j}}{z_{ij}}+O(\ell^0)\,,
\end{equation}
which is proportional to the Carrollian scattering equation $\mathbf{E}_i$, missing only an overall factor of $\partial_{u_i}$. One might worry that, as these Carrollian scattering equations are actually differential operators, we should not infer too much from this fact (e.g., the missing $\partial_{u_i}$ is a differential operator, not an algebraic factor). However, we can simply Fourier transform the entire expression back to momentum space, at which point $\partial_{u_i}$ becomes the multiplicative factor $-\im\,\omega_i$ and the flat limit of $\epsilon_i\cdot\mathcal{T}_i$ is proportional to the algebraic (momentum space) scattering equation $E_i$. Hence, the leading flat limit of \eqref{gluonholcorr} appears to vanish for all $n\geq4$.

It should be noted that \eqref{gluonholcorr} can also be evaluated for $n=2,\ 3$, where there are no scattering equations and the previous arguments do not apply. At two points, none of the differential operators $\epsilon_i\cdot \mathcal{T}_i$ appear and the correlator is (up to a numerical factor):
\begin{equation}
    \mathcal{C}^{\text{hol}}_2=\frac{1}{(x_{12}+\frac{u_{12}}{2\,\ell})^2}\,.
\end{equation}
By demanding that the flat space limit produces an electric branch Carrollian 2-point amplitude (cf., \cite{Alday:2024yyj}), one finds that 
\begin{equation}
    \mathcal{C}^{\text{hol}}_2\propto\frac{ \delta(x_{12})\, \ell}{u_{12}}+O(\ell^0)\,.
\end{equation}
So in order for the flat limit to be finite, one must normalise the Carrollian operators by a factor of $\ell^{-1/2}$ with respect to the AdS ones. At three-points, the correlator reads:
\begin{equation}\label{AdS33pt}
    \mathcal{C}^{\text{hol}}_3\propto \frac{1}{(x_{12}+\frac{u_{12}}{2\ell})(x_{23}+\frac{u_{23}}{2\ell})(x_{31}+\frac{u_{31}}{2\ell})}\,.
\end{equation}
Since the three-point momentum kinematics will contain delta functions $\delta(x_{12})\,\delta(x_{13})$, it follows that the flat limit of \eqref{AdS33pt} vanishes. 

Thus, it seems that for all $n\geq3$ the leading flat limit of the holomorphic gluon boundary correlators \eqref{gluonholcorr} vanishes. Perhaps, in light of the known subtleties associated with flat limits of spinning correlators, there is some more interesting structure lurking at subleading orders in the $\ell\to\infty$ limit. However, one typically finds structures incompatible with Poincar\'e invariance, indicating that the subleading contributions are not associated with flat space, and rather `remember' the fact that they came from AdS. For instance, the subleading contribution to the operator $\epsilon_i\cdot \mathcal{T}_i$ is:
\begin{equation}
    \epsilon_i\cdot \mathcal{T}_i\Big|_{\ell^0}=\sum_{j\neq i}\frac{x_{ij}\,u_{ij}}{z_{ij}}\partial_{u_j}-\frac{x_{ij}\,\Delta_j}{z_{ij}}+\frac{x_{ij}^2\,\partial_{x_j}}{2\,z_{ij}}\,.
\end{equation}
While this is not proportional to the scattering equations, the final term here has derivatives with respect to the celestial circle coordinates $x_j$. Upon Fourier transforming back to momentum space, these will act on any momentum conserving delta functions, breaking momentum conservation and thus Lorentz invariance.

\medskip

Let us conclude with a curious observation. The choice of holomorphic gluon polarizations was, in fact, somewhat subtle in the first place. In~\cite{Roehrig:2020kck}, these were actually motivated by the polarization data for \emph{gravitons} in AdS$_3$, which can naturally be decomposed into boundary holomorphic/anti-holomorphic modes, the former taking the form $\epsilon_{(a}\epsilon_{b)}$. One can show, following identical arguments, that the leading flat space limit of the graviton boundary correlators given in~\cite{Roehrig:2020kck} also vanishes, but this is to be expected since gravity in 3d has no local degrees of freedom. The same is certainly not true of Yang-Mills theory, though, which has a single on-shell polarization and non-vanishing scattering amplitudes.

While~\cite{Roehrig:2020kck} showed that gluon vertex operators with the holomorphic polarization vector \eqref{AdSPol} are in the BRST cohomology of the relevant ambitwistor string, one could wonder if this choice somehow singles out the (trivial) Chern-Simons bulk dynamics of the conjectured Yang-Mills-Chern-Simons theory in the flat limit. Perhaps an alternative choice of polarization could lead to better results.


Consider, for instance, an alternative polarization vector in AdS$_3$:
\begin{equation}\label{neweps}
    \epsilon'_a=\langle \lam|\mathrm{t}_a|\iota\rangle\,,\qquad |\iota\rangle=(1,0)\,,
\end{equation}
a choice which is loosely motivated by the structure of the on-shell gluon polarization vector in flat space. One can quickly show that the associated gluon vertex operator in the ambitwistor string is \emph{not} BRST-closed, indicating that this should not be an admissible choice of vertex operator in the theory.
Nevertheless, we can follow the example of the scalar CHY formula in AdS (which was computed in a wildly anomalous ambitwistor string model yet gave a correct result) and compute the correlator of holomorphic gluons with this new polarization vector. The main difference from \eqref{gluonholcorr} is that the leading order contribution to $\epsilon_i\cdot\mathcal{T}_i$ becomes:
\begin{equation}
    \epsilon'_i\cdot \mathcal{T}_i= 2\ell\, \sum_{j\neq i}\frac{x_{ij}}{z_{ij}}\,\partial_{u_j}+O(\ell^0)\,,
\end{equation}
which is \emph{not} proportional to a Carrollian scattering equation. Consequently, there is no reason for the flat space limit of the correlator to vanish.

This can be confirmed at four points, where, after carrying out the moduli space integral and Fourier transforming to momentum space, the color-ordered amplitude reads:
\begin{equation}\label{flat4pt}
    A_{4,\text{flat}}(1,2,3,4)\propto \delta^3\!\left(\sum_{i=1}^4 k_i\right)\, \frac{x_{13}\,x_{24}}{x_{12}\,x_{34}}\,.
\end{equation}
While non-zero, this is \emph{not} the four-point amplitude of Yang-Mills theory in 3-dimensions. Indeed, the 4-point gluon color-ordered amplitude in 3d is:
\begin{equation}
    A_{4,\text{YM}}(1,2,3,4)\propto  \delta^3\!\left(\sum_{i=1}^4 k_i\right) \left(\,\frac{x_{31}}{x_{12}}+ \frac{x_{12}}{x_{32}}+\, \frac{x_{12}\,x_{13}}{x_{23}\,x_{14}} +\, \frac{x_{13}\,x_{32}}{x_{34}\,x_{12}}\right).
\end{equation}
The amplitude \eqref{flat4pt} \emph{can} be obtained as the dimensional reduction from 4d to 3d of a certain scalar operator in 4d $\mathcal{N}=4$ super-Yang-Mills, though. 

Starting from the on-shell 4d superfield:
\begin{equation}
    \Omega=g^+ +\eta_A \lambda^A +\frac{1}{2}\eta_A\eta_B S^{AB} +\frac{1}{3!}\eta_A\eta_B\eta_C\bar{\lambda}^{ABC}+\eta_1\eta_2\eta_3\eta_4 g^- \, ,
\end{equation}
where $S^{AB}$ are the six scalars, then one can consider the amplitude :
\begin{equation}
    A_{4}^{D=4}(1^{S^{34}},2^{S^{12}}, 3^{S^{14}},4^{S^{32}}) \propto \delta^4\!\left(\sum_{i=1}^4 k_i\right) \frac{\langle 13\rangle\langle 24\rangle}{\langle12 \rangle \langle 34\rangle} \, \,,
\end{equation}
where $\la i\,j\ra=\kappa_i^{\alpha}\,\kappa_{j}^{\beta}\,\epsilon_{\beta\alpha}$ are the usual spinor helicity invariants constructed from on-shell 4-momenta $k_i^{\alpha\dot\alpha}=\kappa_i^{\alpha}\,\tilde{\kappa}_{i}^{\dot\alpha}$. Remarkably, dimensional reduction of this amplitude to $D=3$ gives precisely \eqref{flat4pt}. 

It is curious that the choice of polarization \eqref{neweps} seems to pick out a scalar amplitude in the flat limit. One might take this as proof that the lack of BRST-closure for the associated vertex operators rendered the entire calculation meaningless from the start, but the fact that the resulting flat space amplitude \eqref{flat4pt} is a dimensional reduction of a supersymmetric \emph{Yang-Mills} amplitude is certainly tantalizing. Clearly, the relationship between different polarizations and the flat limits of formula \eqref{gluonholcorr} requires a more systematic treatment!


\section{Discussion} \label{sec:6}

The study of Carrollian scattering amplitudes and their relation to the flat-space limit of AdS correlators remains an evolving field. In this work, we employed the CHY formalism and ambitwistor strings to derive all-multiplicity formulae for Carrollian amplitudes in bi-adjoint scalar theory, Yang–Mills theory, and NS–NS gravity. We also demonstrated explicitly how the scalar Carrollian  CHY formula emerges from the flat-space limit of the corresponding AdS CHY representation.

There are several promising directions for future research. First, CHY formulae apply to a wide range of quantum field theories, including Einstein–Yang–Mills, Born–Infeld, and nonlinear sigma models~\cite{Cachazo:2014xea}. It would be valuable to understand their Carrollian counterparts. In parallel, worldsheet models with null infinity as their target space reproduce four-dimensional Yang–Mills and gravity amplitudes~\cite{Adamo:2014yya,Adamo:2015fwa}. Clarifying their connection to Carrollian CHY formulae may provide insights into the intrinsic structure of the dual Carrollian CFT.

It would be very interesting to reproduce the Carrollian CHY formulae for the spinning case in general spacetime dimensions from the flat space limit of AdS correlators. As we discussed in section \ref{sec:5}, a systematic study of the flat space limit of spinning correlators is required, and surely AdS$_3$ is a natural first proving ground. This is particularly true as it is the setting of the only known CHY formulae for spinning AdS correlators.

Most of the work on Carrollian amplitudes is restricted to the case where external particles are massless. The proper definition for massive Carrollian amplitudes is still absent, compared to its celestial counterpart~\cite{Pasterski:2016qvg,Pasterski:2017kqt}. It would be interesting to study how to modify the prescription of taking the flat space limit of AdS correlators used in this work so that massive amplitudes are reproduced. This would require a different scaling where the conformal dimensions are not fixed, but taken to scale with the AdS radius, with the ratio between the two being proportional to the mass.

It has been shown that the CHY formulae have a nice interpretation from twisted intersection theory which manifests the double copy~\cite{Mizera:2017cqs,Mizera:2019gea,Mizera:2019blq}. The relevant operator-valued twisted forms for celestial CHY formulae were constructed in~\cite{Casali:2020uvr}, and it would be interesting to construct them for Carrollian CHY formulae to study the double copy for Carrollian amplitudes. 

Finally, it would be useful to solve the Carrollian scattering equations directly. Then one can replace the operator-valued equations with c-numbers. This might be helpful in setting up a boostrap procedure for Carrollian amplitudes. We also expect a close connection between the Carrollian scattering equations and the differential quations for Carrollian amplitudes
obtained in~\cite{Ruzziconi:2024zkr}. We leave these various avenues for future work.


\section*{Acknowledgements}
We would like to thank Kai R\"ohrig, Sean Seet, Joan S\'imon and David Skinner for useful conversations. TA is supported by a Royal Society University Research Fellowship, the Simons Collaboration on Celestial Holography CH-00001550-11, the ERC Consolidator/UKRI Frontier grant TwistorQFT EP/Z000157/1 and the STFC consolidated grant ST/X000494/1. IS is supported by an STFC studentship. BZ is supported by the Simons Collaboration on Celestial Holography CH-00001550-11.

\bibliographystyle{JHEP}
\bibliography{cope.bib}

\providecommand{\href}[2]{#2}\begingroup\raggedright\begin{thebibliography}{100}

\bibitem{Strominger:2017zoo}
A.~Strominger, {\it {Lectures on the Infrared Structure of Gravity and Gauge Theory}},  \href{http://arxiv.org/abs/1703.05448}{{\tt arXiv:1703.05448}}.

\bibitem{Raclariu:2021zjz}
A.-M. Raclariu, {\it {Lectures on Celestial Holography}},  \href{http://arxiv.org/abs/2107.02075}{{\tt arXiv:2107.02075}}.

\bibitem{Pasterski:2021raf}
S.~Pasterski, M.~Pate, and A.-M. Raclariu, {\it {Celestial Holography}},  in {\em {Snowmass 2021}}, 11, 2021.
\newblock \href{http://arxiv.org/abs/2111.11392}{{\tt arXiv:2111.11392}}.

\bibitem{Donnay:2023mrd}
L.~Donnay, {\it {Celestial holography: An asymptotic symmetry perspective}},  {\em Phys. Rept.} {\bf 1073} (2024) 1--41, [\href{http://arxiv.org/abs/2310.12922}{{\tt arXiv:2310.12922}}].

\bibitem{Donnay:2022aba}
L.~Donnay, A.~Fiorucci, Y.~Herfray, and R.~Ruzziconi, {\it {Carrollian Perspective on Celestial Holography}},  {\em Phys. Rev. Lett.} {\bf 129} (2022), no.~7 071602, [\href{http://arxiv.org/abs/2202.04702}{{\tt arXiv:2202.04702}}].

\bibitem{Bagchi:2022emh}
A.~Bagchi, S.~Banerjee, R.~Basu, and S.~Dutta, {\it {Scattering Amplitudes: Celestial and Carrollian}},  {\em Phys. Rev. Lett.} {\bf 128} (2022), no.~24 241601, [\href{http://arxiv.org/abs/2202.08438}{{\tt arXiv:2202.08438}}].

\bibitem{Donnay:2022wvx}
L.~Donnay, A.~Fiorucci, Y.~Herfray, and R.~Ruzziconi, {\it {Bridging Carrollian and celestial holography}},  {\em Phys. Rev. D} {\bf 107} (2023), no.~12 126027, [\href{http://arxiv.org/abs/2212.12553}{{\tt arXiv:2212.12553}}].

\bibitem{Bagchi:2025vri}
A.~Bagchi, A.~Banerjee, P.~Dhivakar, S.~Mondal, and A.~Shukla, {\it {The Carrollian Kaleidoscope}},  \href{http://arxiv.org/abs/2506.16164}{{\tt arXiv:2506.16164}}.

\bibitem{Nguyen:2025zhg}
K.~Nguyen, {\it {Lectures on Carrollian Holography}},  \href{http://arxiv.org/abs/2511.10162}{{\tt arXiv:2511.10162}}.

\bibitem{Maldacena:1997re}
J.~M. Maldacena, {\it {The Large $N$ limit of superconformal field theories and supergravity}},  {\em Adv. Theor. Math. Phys.} {\bf 2} (1998) 231--252, [\href{http://arxiv.org/abs/hep-th/9711200}{{\tt hep-th/9711200}}].

\bibitem{Witten:1998qj}
E.~Witten, {\it {Anti de Sitter space and holography}},  {\em Adv. Theor. Math. Phys.} {\bf 2} (1998) 253--291, [\href{http://arxiv.org/abs/hep-th/9802150}{{\tt hep-th/9802150}}].

\bibitem{Gubser:1998bc}
S.~S. Gubser, I.~R. Klebanov, and A.~M. Polyakov, {\it {Gauge theory correlators from noncritical string theory}},  {\em Phys. Lett. B} {\bf 428} (1998) 105--114, [\href{http://arxiv.org/abs/hep-th/9802109}{{\tt hep-th/9802109}}].

\bibitem{Susskind:1998vk}
L.~Susskind, {\it {Holography in the flat space limit}},  {\em AIP Conf. Proc.} {\bf 493} (1999), no.~1 98--112, [\href{http://arxiv.org/abs/hep-th/9901079}{{\tt hep-th/9901079}}].

\bibitem{Polchinski:1999ry}
J.~Polchinski, {\it {S matrices from AdS space-time}},  \href{http://arxiv.org/abs/hep-th/9901076}{{\tt hep-th/9901076}}.

\bibitem{Giddings:1999jq}
S.~B. Giddings, {\it {Flat space scattering and bulk locality in the AdS / CFT correspondence}},  {\em Phys. Rev. D} {\bf 61} (2000) 106008, [\href{http://arxiv.org/abs/hep-th/9907129}{{\tt hep-th/9907129}}].

\bibitem{Giddings:1999qu}
S.~B. Giddings, {\it {The Boundary S matrix and the AdS to CFT dictionary}},  {\em Phys. Rev. Lett.} {\bf 83} (1999) 2707--2710, [\href{http://arxiv.org/abs/hep-th/9903048}{{\tt hep-th/9903048}}].

\bibitem{Maldacena:2002vr}
J.~M. Maldacena, {\it {Non-Gaussian features of primordial fluctuations in single field inflationary models}},  {\em JHEP} {\bf 05} (2003) 013, [\href{http://arxiv.org/abs/astro-ph/0210603}{{\tt astro-ph/0210603}}].

\bibitem{Raju:2012zr}
S.~Raju, {\it {New Recursion Relations and a Flat Space Limit for AdS/CFT Correlators}},  {\em Phys. Rev. D} {\bf 85} (2012) 126009, [\href{http://arxiv.org/abs/1201.6449}{{\tt arXiv:1201.6449}}].

\bibitem{Maldacena:2011nz}
J.~M. Maldacena and G.~L. Pimentel, {\it {On graviton non-Gaussianities during inflation}},  {\em JHEP} {\bf 09} (2011) 045, [\href{http://arxiv.org/abs/1104.2846}{{\tt arXiv:1104.2846}}].

\bibitem{Fitzpatrick:2011dm}
A.~L. Fitzpatrick and J.~Kaplan, {\it {Unitarity and the Holographic S-Matrix}},  {\em JHEP} {\bf 10} (2012) 032, [\href{http://arxiv.org/abs/1112.4845}{{\tt arXiv:1112.4845}}].

\bibitem{Paulos:2016fap}
M.~F. Paulos, J.~Penedones, J.~Toledo, B.~C. van Rees, and P.~Vieira, {\it {The S-matrix bootstrap. Part I: QFT in AdS}},  {\em JHEP} {\bf 11} (2017) 133, [\href{http://arxiv.org/abs/1607.06109}{{\tt arXiv:1607.06109}}].

\bibitem{Alday:2017vkk}
L.~F. Alday and S.~Caron-Huot, {\it {Gravitational S-matrix from CFT dispersion relations}},  {\em JHEP} {\bf 12} (2018) 017, [\href{http://arxiv.org/abs/1711.02031}{{\tt arXiv:1711.02031}}].

\bibitem{Hijano:2019qmi}
E.~Hijano, {\it {Flat space physics from AdS/CFT}},  {\em JHEP} {\bf 07} (2019) 132, [\href{http://arxiv.org/abs/1905.02729}{{\tt arXiv:1905.02729}}].

\bibitem{Komatsu:2020sag}
S.~Komatsu, M.~F. Paulos, B.~C. Van~Rees, and X.~Zhao, {\it {Landau diagrams in AdS and S-matrices from conformal correlators}},  {\em JHEP} {\bf 11} (2020) 046, [\href{http://arxiv.org/abs/2007.13745}{{\tt arXiv:2007.13745}}].

\bibitem{vanRees:2022zmr}
B.~C. van Rees and X.~Zhao, {\it {Quantum Field Theory in AdS Space instead of Lehmann-Symanzik-Zimmerman Axioms}},  {\em Phys. Rev. Lett.} {\bf 130} (2023), no.~19 191601, [\href{http://arxiv.org/abs/2210.15683}{{\tt arXiv:2210.15683}}].

\bibitem{Penedones:2010ue}
J.~Penedones, {\it {Writing CFT correlation functions as AdS scattering amplitudes}},  {\em JHEP} {\bf 03} (2011) 025, [\href{http://arxiv.org/abs/1011.1485}{{\tt arXiv:1011.1485}}].

\bibitem{Fitzpatrick:2011ia}
A.~L. Fitzpatrick, J.~Kaplan, J.~Penedones, S.~Raju, and B.~C. van Rees, {\it {A Natural Language for AdS/CFT Correlators}},  {\em JHEP} {\bf 11} (2011) 095, [\href{http://arxiv.org/abs/1107.1499}{{\tt arXiv:1107.1499}}].

\bibitem{Paulos:2011ie}
M.~F. Paulos, {\it {Towards Feynman rules for Mellin amplitudes}},  {\em JHEP} {\bf 10} (2011) 074, [\href{http://arxiv.org/abs/1107.1504}{{\tt arXiv:1107.1504}}].

\bibitem{Bagchi:2023cen}
A.~Bagchi, P.~Dhivakar, and S.~Dutta, {\it {Holography in flat spacetimes: the case for Carroll}},  {\em JHEP} {\bf 08} (2024) 144, [\href{http://arxiv.org/abs/2311.11246}{{\tt arXiv:2311.11246}}].

\bibitem{Alday:2024yyj}
L.~F. Alday, M.~Nocchi, R.~Ruzziconi, and A.~Yelleshpur~Srikant, {\it {Carrollian amplitudes from holographic correlators}},  {\em JHEP} {\bf 03} (2025) 158, [\href{http://arxiv.org/abs/2406.19343}{{\tt arXiv:2406.19343}}].

\bibitem{Berenstein:2025tts}
D.~Berenstein and J.~Simon, {\it {Aspects of the bulk flat space limit in AdS/CFT}},  \href{http://arxiv.org/abs/2510.23697}{{\tt arXiv:2510.23697}}.

\bibitem{Bagchi:2023fbj}
A.~Bagchi, P.~Dhivakar, and S.~Dutta, {\it {AdS Witten diagrams to Carrollian correlators}},  {\em JHEP} {\bf 04} (2023) 135, [\href{http://arxiv.org/abs/2303.07388}{{\tt arXiv:2303.07388}}].

\bibitem{Salzer:2023jqv}
J.~Salzer, {\it {An embedding space approach to Carrollian CFT correlators for flat space holography}},  {\em JHEP} {\bf 10} (2023) 084, [\href{http://arxiv.org/abs/2304.08292}{{\tt arXiv:2304.08292}}].

\bibitem{Saha:2023abr}
A.~Saha, {\it {w$_{1+\infty}$ and Carrollian holography}},  {\em JHEP} {\bf 05} (2024) 145, [\href{http://arxiv.org/abs/2308.03673}{{\tt arXiv:2308.03673}}].

\bibitem{Nguyen:2023vfz}
K.~Nguyen and P.~West, {\it {Carrollian Conformal Fields and Flat Holography}},  {\em Universe} {\bf 9} (2023), no.~9 385, [\href{http://arxiv.org/abs/2305.02884}{{\tt arXiv:2305.02884}}].

\bibitem{Nguyen:2023miw}
K.~Nguyen, {\it {Carrollian conformal correlators and massless scattering amplitudes}},  {\em JHEP} {\bf 01} (2024) 076, [\href{http://arxiv.org/abs/2311.09869}{{\tt arXiv:2311.09869}}].

\bibitem{Mason:2023mti}
L.~Mason, R.~Ruzziconi, and A.~Yelleshpur~Srikant, {\it {Carrollian amplitudes and celestial symmetries}},  {\em JHEP} {\bf 05} (2024) 012, [\href{http://arxiv.org/abs/2312.10138}{{\tt arXiv:2312.10138}}].

\bibitem{Liu:2024nfc}
W.-B. Liu, J.~Long, and X.-Q. Ye, {\it {Feynman rules and loop structure of Carrollian amplitudes}},  {\em JHEP} {\bf 05} (2024) 213, [\href{http://arxiv.org/abs/2402.04120}{{\tt arXiv:2402.04120}}].

\bibitem{Have:2024dff}
E.~Have, K.~Nguyen, S.~Prohazka, and J.~Salzer, {\it {Massive carrollian fields at timelike infinity}},  {\em JHEP} {\bf 07} (2024) 054, [\href{http://arxiv.org/abs/2402.05190}{{\tt arXiv:2402.05190}}].

\bibitem{Stieberger:2024shv}
S.~Stieberger, T.~R. Taylor, and B.~Zhu, {\it {Carrollian Amplitudes from Strings}},  {\em JHEP} {\bf 04} (2024) 127, [\href{http://arxiv.org/abs/2402.14062}{{\tt arXiv:2402.14062}}].

\bibitem{Adamo:2024mqn}
T.~Adamo, W.~Bu, P.~Tourkine, and B.~Zhu, {\it {Eikonal amplitudes on the celestial sphere}},  {\em JHEP} {\bf 10} (2024) 192, [\href{http://arxiv.org/abs/2405.15594}{{\tt arXiv:2405.15594}}].

\bibitem{Banerjee:2024hvb}
S.~Banerjee, R.~Basu, and S.~Atul~Bhatkar, {\it {Light transformation: a celestial and Carrollian perspective}},  {\em JHEP} {\bf 12} (2024) 122, [\href{http://arxiv.org/abs/2407.08379}{{\tt arXiv:2407.08379}}].

\bibitem{Ruzziconi:2024zkr}
R.~Ruzziconi, S.~Stieberger, T.~R. Taylor, and B.~Zhu, {\it {Differential equations for Carrollian amplitudes}},  {\em JHEP} {\bf 09} (2024) 149, [\href{http://arxiv.org/abs/2407.04789}{{\tt arXiv:2407.04789}}].

\bibitem{Ruzziconi:2024kzo}
R.~Ruzziconi and A.~Saha, {\it {Holographic Carrollian currents for massless scattering}},  {\em JHEP} {\bf 01} (2025) 169, [\href{http://arxiv.org/abs/2411.04902}{{\tt arXiv:2411.04902}}].

\bibitem{Chakrabortty:2024bvm}
S.~Chakrabortty, S.~Hegde, and A.~Maurya, {\it {Differential representation for Carrollian correlators}},  {\em JHEP} {\bf 08} (2025) 126, [\href{http://arxiv.org/abs/2411.09641}{{\tt arXiv:2411.09641}}].

\bibitem{Nguyen:2025sqk}
K.~Nguyen and J.~Salzer, {\it {Operator product expansion in Carrollian CFT}},  {\em JHEP} {\bf 07} (2025) 193, [\href{http://arxiv.org/abs/2503.15607}{{\tt arXiv:2503.15607}}].

\bibitem{Lipstein:2025jfj}
A.~Lipstein, R.~Ruzziconi, and A.~Yelleshpur~Srikant, {\it {Towards a flat space Carrollian hologram from AdS$_{4}$/CFT$_{3}$}},  {\em JHEP} {\bf 06} (2025) 073, [\href{http://arxiv.org/abs/2504.10291}{{\tt arXiv:2504.10291}}].

\bibitem{deGioia:2025mwt}
L.~P. de~Gioia and A.-M. Raclariu, {\it {Infinite towers of 2d symmetry algebras from Carrollian limit of 3d CFT}},  \href{http://arxiv.org/abs/2508.19981}{{\tt arXiv:2508.19981}}.

\bibitem{Surubaru:2025fmg}
I.~Surubaru and B.~Zhu, {\it {Carrollian amplitudes and holographic correlators in AdS3/CFT2}},  {\em Phys. Rev. D} {\bf 112} (2025), no.~2 026023, [\href{http://arxiv.org/abs/2504.07650}{{\tt arXiv:2504.07650}}].

\bibitem{Liu:2024llk}
W.-B. Liu, J.~Long, H.-Y. Xiao, and J.-L. Yang, {\it {On the definition of Carrollian amplitudes in general dimensions}},  {\em JHEP} {\bf 11} (2024) 027, [\href{http://arxiv.org/abs/2407.20816}{{\tt arXiv:2407.20816}}].

\bibitem{Kulkarni:2025qcx}
H.~Kulkarni, R.~Ruzziconi, and A.~Yelleshpur~Srikant, {\it {On Carrollian and Celestial Correlators in General Dimensions}},  \href{http://arxiv.org/abs/2508.06602}{{\tt arXiv:2508.06602}}.

\bibitem{Cachazo:2013hca}
F.~Cachazo, S.~He, and E.~Y. Yuan, {\it {Scattering of Massless Particles in Arbitrary Dimensions}},  {\em Phys. Rev. Lett.} {\bf 113} (2014), no.~17 171601, [\href{http://arxiv.org/abs/1307.2199}{{\tt arXiv:1307.2199}}].

\bibitem{Cachazo:2014xea}
F.~Cachazo, S.~He, and E.~Y. Yuan, {\it {Scattering Equations and Matrices: From Einstein To Yang-Mills, DBI and NLSM}},  {\em JHEP} {\bf 07} (2015) 149, [\href{http://arxiv.org/abs/1412.3479}{{\tt arXiv:1412.3479}}].

\bibitem{Mason:2013sva}
L.~Mason and D.~Skinner, {\it {Ambitwistor strings and the scattering equations}},  {\em JHEP} {\bf 07} (2014) 048, [\href{http://arxiv.org/abs/1311.2564}{{\tt arXiv:1311.2564}}].

\bibitem{Eberhardt:2020ewh}
L.~Eberhardt, S.~Komatsu, and S.~Mizera, {\it {Scattering equations in AdS: scalar correlators in arbitrary dimensions}},  {\em JHEP} {\bf 11} (2020) 158, [\href{http://arxiv.org/abs/2007.06574}{{\tt arXiv:2007.06574}}].

\bibitem{Roehrig:2020kck}
K.~Roehrig and D.~Skinner, {\it {Ambitwistor strings and the scattering equations on AdS$_{3}${\texttimes}S$^{3}$}},  {\em JHEP} {\bf 02} (2022) 073, [\href{http://arxiv.org/abs/2007.07234}{{\tt arXiv:2007.07234}}].

\bibitem{Li:2021snj}
Y.-Z. Li, {\it {Notes on flat-space limit of AdS/CFT}},  {\em JHEP} {\bf 09} (2021) 027, [\href{http://arxiv.org/abs/2106.04606}{{\tt arXiv:2106.04606}}].

\bibitem{Berenstein:2025qhb}
D.~Berenstein and Z.~Li, {\it {Spinning Fields in Lorentzian AdS}},  \href{http://arxiv.org/abs/2511.15780}{{\tt arXiv:2511.15780}}.

\bibitem{Cachazo:2013iea}
F.~Cachazo, S.~He, and E.~Y. Yuan, {\it {Scattering of Massless Particles: Scalars, Gluons and Gravitons}},  {\em JHEP} {\bf 07} (2014) 033, [\href{http://arxiv.org/abs/1309.0885}{{\tt arXiv:1309.0885}}].

\bibitem{Cachazo:2014nsa}
F.~Cachazo, S.~He, and E.~Y. Yuan, {\it {Einstein-Yang-Mills Scattering Amplitudes From Scattering Equations}},  {\em JHEP} {\bf 01} (2015) 121, [\href{http://arxiv.org/abs/1409.8256}{{\tt arXiv:1409.8256}}].

\bibitem{Berkovits:2013xba}
N.~Berkovits, {\it {Infinite Tension Limit of the Pure Spinor Superstring}},  {\em JHEP} {\bf 03} (2014) 017, [\href{http://arxiv.org/abs/1311.4156}{{\tt arXiv:1311.4156}}].

\bibitem{Adamo:2013tsa}
T.~Adamo, E.~Casali, and D.~Skinner, {\it {Ambitwistor strings and the scattering equations at one loop}},  {\em JHEP} {\bf 04} (2014) 104, [\href{http://arxiv.org/abs/1312.3828}{{\tt arXiv:1312.3828}}].

\bibitem{Pasterski:2016qvg}
S.~Pasterski, S.-H. Shao, and A.~Strominger, {\it {Flat Space Amplitudes and Conformal Symmetry of the Celestial Sphere}},  {\em Phys. Rev. D} {\bf 96} (2017), no.~6 065026, [\href{http://arxiv.org/abs/1701.00049}{{\tt arXiv:1701.00049}}].

\bibitem{Pasterski:2017kqt}
S.~Pasterski and S.-H. Shao, {\it {Conformal basis for flat space amplitudes}},  {\em Phys. Rev. D} {\bf 96} (2017), no.~6 065022, [\href{http://arxiv.org/abs/1705.01027}{{\tt arXiv:1705.01027}}].

\bibitem{Adamo:2019ipt}
T.~Adamo, L.~Mason, and A.~Sharma, {\it {Celestial amplitudes and conformal soft theorems}},  {\em Class. Quant. Grav.} {\bf 36} (2019), no.~20 205018, [\href{http://arxiv.org/abs/1905.09224}{{\tt arXiv:1905.09224}}].

\bibitem{Casali:2020uvr}
E.~Casali and A.~Sharma, {\it {Celestial double copy from the worldsheet}},  {\em JHEP} {\bf 05} (2021) 157, [\href{http://arxiv.org/abs/2011.10052}{{\tt arXiv:2011.10052}}].

\bibitem{Banerjee:2018fgd}
S.~Banerjee, {\it {Symmetries of free massless particles and soft theorems}},  {\em Gen. Rel. Grav.} {\bf 51} (2019), no.~9 128, [\href{http://arxiv.org/abs/1804.06646}{{\tt arXiv:1804.06646}}].

\bibitem{Banerjee:2018gce}
S.~Banerjee, {\it {Null Infinity and Unitary Representation of The Poincare Group}},  {\em JHEP} {\bf 01} (2019) 205, [\href{http://arxiv.org/abs/1801.10171}{{\tt arXiv:1801.10171}}].

\bibitem{Pasterski:2021dqe}
S.~Pasterski, A.~Puhm, and E.~Trevisani, {\it {Revisiting the conformally soft sector with celestial diamonds}},  {\em JHEP} {\bf 11} (2021) 143, [\href{http://arxiv.org/abs/2105.09792}{{\tt arXiv:2105.09792}}].

\bibitem{Fairlie:1972zz}
D.~B. Fairlie and D.~E. Roberts, {\it {Dual models without tachyons - A new approach}}, .

\bibitem{Fairlie:2008dg}
D.~B. Fairlie, {\it {A Coding of Real Null Four-Momenta into World-Sheet Coordinates}},  {\em Adv. Math. Phys.} {\bf 2009} (2009) 284689, [\href{http://arxiv.org/abs/0805.2263}{{\tt arXiv:0805.2263}}].

\bibitem{Cachazo:2013gna}
F.~Cachazo, S.~He, and E.~Y. Yuan, {\it {Scattering equations and Kawai-Lewellen-Tye orthogonality}},  {\em Phys. Rev. D} {\bf 90} (2014), no.~6 065001, [\href{http://arxiv.org/abs/1306.6575}{{\tt arXiv:1306.6575}}].

\bibitem{Dolan:2014ega}
L.~Dolan and P.~Goddard, {\it {The Polynomial Form of the Scattering Equations}},  {\em JHEP} {\bf 07} (2014) 029, [\href{http://arxiv.org/abs/1402.7374}{{\tt arXiv:1402.7374}}].

\bibitem{Dolan:2015iln}
L.~Dolan and P.~Goddard, {\it {General Solution of the Scattering Equations}},  {\em JHEP} {\bf 10} (2016) 149, [\href{http://arxiv.org/abs/1511.09441}{{\tt arXiv:1511.09441}}].

\bibitem{Gross:1987ar}
D.~J. Gross and P.~F. Mende, {\it {String Theory Beyond the Planck Scale}},  {\em Nucl. Phys. B} {\bf 303} (1988) 407--454.

\bibitem{Dolan:2013isa}
L.~Dolan and P.~Goddard, {\it {Proof of the Formula of Cachazo, He and Yuan for Yang-Mills Tree Amplitudes in Arbitrary Dimension}},  {\em JHEP} {\bf 05} (2014) 010, [\href{http://arxiv.org/abs/1311.5200}{{\tt arXiv:1311.5200}}].

\bibitem{Casali:2015vta}
E.~Casali, Y.~Geyer, L.~Mason, R.~Monteiro, and K.~A. Roehrig, {\it {New Ambitwistor String Theories}},  {\em JHEP} {\bf 11} (2015) 038, [\href{http://arxiv.org/abs/1506.08771}{{\tt arXiv:1506.08771}}].

\bibitem{Isenberg:1978kk}
J.~Isenberg, P.~B. Yasskin, and P.~S. Green, {\it {Nonselfdual Gauge Fields}},  {\em Phys. Lett. B} {\bf 78} (1978) 462--464.

\bibitem{Witten:1978xx}
E.~Witten, {\it {An Interpretation of Classical Yang-Mills Theory}},  {\em Phys. Lett. B} {\bf 77} (1978) 394--398.

\bibitem{LeBrun:1983}
C.~LeBrun, {\it {Spaces of Complex Null Geodesics in Complex Riemannian Geometry}},  {\em Trans. Amer. Math. Soc.} {\bf 278} (1983) 209--231.

\bibitem{Witten:1985nt}
E.~Witten, {\it {Twistor - Like Transform in Ten-Dimensions}},  {\em Nucl. Phys. B} {\bf 266} (1986) 245--264.

\bibitem{Berkovits:2004jj}
N.~Berkovits and E.~Witten, {\it {Conformal supergravity in twistor-string theory}},  {\em JHEP} {\bf 08} (2004) 009, [\href{http://arxiv.org/abs/hep-th/0406051}{{\tt hep-th/0406051}}].

\bibitem{Azevedo:2017lkz}
T.~Azevedo and O.~T. Engelund, {\it {Ambitwistor formulations of R$^{2}$ gravity and (DF)$^{2}$ gauge theories}},  {\em JHEP} {\bf 11} (2017) 052, [\href{http://arxiv.org/abs/1707.02192}{{\tt arXiv:1707.02192}}].

\bibitem{Adamo:2018hzd}
T.~Adamo, E.~Casali, and S.~Nekovar, {\it {Yang-Mills theory from the worldsheet}},  {\em Phys. Rev. D} {\bf 98} (2018), no.~8 086022, [\href{http://arxiv.org/abs/1807.09171}{{\tt arXiv:1807.09171}}].

\bibitem{Seet:2025mes}
S.~Seet, {\it {Single-trace current correlators for 2d models of 4d gluon scattering}},  \href{http://arxiv.org/abs/2509.12200}{{\tt arXiv:2509.12200}}.

\bibitem{Berkovits:2018jvm}
N.~Berkovits and M.~Lize, {\it {Field theory actions for ambitwistor string and superstring}},  {\em JHEP} {\bf 09} (2018) 097, [\href{http://arxiv.org/abs/1807.07661}{{\tt arXiv:1807.07661}}].

\bibitem{Figueroa-OFarrill:2024wgs}
J.~M. Figueroa-O'Farrill and G.~S. Vishwa, {\it {The BRST quantisation of chiral BMS-like field theories}},  {\em J. Math. Phys.} {\bf 66} (2025), no.~4 042303, [\href{http://arxiv.org/abs/2407.12778}{{\tt arXiv:2407.12778}}].

\bibitem{Adamo:2014wea}
T.~Adamo, E.~Casali, and D.~Skinner, {\it {A Worldsheet Theory for Supergravity}},  {\em JHEP} {\bf 02} (2015) 116, [\href{http://arxiv.org/abs/1409.5656}{{\tt arXiv:1409.5656}}].

\bibitem{Adamo:2018ege}
T.~Adamo, E.~Casali, and S.~Nekovar, {\it {Ambitwistor string vertex operators on curved backgrounds}},  {\em JHEP} {\bf 01} (2019) 213, [\href{http://arxiv.org/abs/1809.04489}{{\tt arXiv:1809.04489}}].

\bibitem{Gomez:2013wza}
H.~Gomez and E.~Y. Yuan, {\it {N-point tree-level scattering amplitude in the new Berkovits` string}},  {\em JHEP} {\bf 04} (2014) 046, [\href{http://arxiv.org/abs/1312.5485}{{\tt arXiv:1312.5485}}].

\bibitem{Adamo:2015hoa}
T.~Adamo and E.~Casali, {\it {Scattering equations, supergravity integrands, and pure spinors}},  {\em JHEP} {\bf 05} (2015) 120, [\href{http://arxiv.org/abs/1502.06826}{{\tt arXiv:1502.06826}}].

\bibitem{Chandia:2015xfa}
O.~Chandia and B.~C. Vallilo, {\it {On-shell type II supergravity from the ambitwistor pure spinor string}},  {\em Class. Quant. Grav.} {\bf 33} (2016), no.~18 185003, [\href{http://arxiv.org/abs/1511.03329}{{\tt arXiv:1511.03329}}].

\bibitem{Gomez:2021qfd}
H.~Gomez, R.~L. Jusinskas, and A.~Lipstein, {\it {Cosmological Scattering Equations}},  {\em Phys. Rev. Lett.} {\bf 127} (2021), no.~25 251604, [\href{http://arxiv.org/abs/2106.11903}{{\tt arXiv:2106.11903}}].

\bibitem{Gomez:2021ujt}
H.~Gomez, R.~Lipinski~Jusinskas, and A.~Lipstein, {\it {Cosmological scattering equations at tree-level and one-loop}},  {\em JHEP} {\bf 07} (2022) 004, [\href{http://arxiv.org/abs/2112.12695}{{\tt arXiv:2112.12695}}].

\bibitem{Armstrong:2022csc}
C.~Armstrong, H.~Gomez, R.~Lipinski~Jusinskas, A.~Lipstein, and J.~Mei, {\it {Effective field theories and cosmological scattering equations}},  {\em JHEP} {\bf 08} (2022) 054, [\href{http://arxiv.org/abs/2204.08931}{{\tt arXiv:2204.08931}}].

\bibitem{Freedman:1998tz}
D.~Z. Freedman, S.~D. Mathur, A.~Matusis, and L.~Rastelli, {\it {Correlation functions in the CFT(d) / AdS(d+1) correspondence}},  {\em Nucl. Phys. B} {\bf 546} (1999) 96--118, [\href{http://arxiv.org/abs/hep-th/9804058}{{\tt hep-th/9804058}}].

\bibitem{Barnich:2012aw}
G.~Barnich, A.~Gomberoff, and H.~A. Gonzalez, {\it {The Flat limit of three dimensional asymptotically anti-de Sitter spacetimes}},  {\em Phys. Rev. D} {\bf 86} (2012) 024020, [\href{http://arxiv.org/abs/1204.3288}{{\tt arXiv:1204.3288}}].

\bibitem{Poole:2018koa}
A.~Poole, K.~Skenderis, and M.~Taylor, {\it {(A)dS$\mathbf{_4}$ in Bondi gauge}},  {\em Class. Quant. Grav.} {\bf 36} (2019), no.~9 095005, [\href{http://arxiv.org/abs/1812.05369}{{\tt arXiv:1812.05369}}].

\bibitem{Compere:2019bua}
G.~Comp{\`e}re, A.~Fiorucci, and R.~Ruzziconi, {\it {The $\Lambda$-BMS$_4$ group of dS$_4$ and new boundary conditions for AdS$_4$}},  {\em Class. Quant. Grav.} {\bf 36} (2019), no.~19 195017, [\href{http://arxiv.org/abs/1905.00971}{{\tt arXiv:1905.00971}}]. [Erratum: Class.Quant.Grav. 38, 229501 (2021)].

\bibitem{Compere:2020lrt}
G.~Comp{\`e}re, A.~Fiorucci, and R.~Ruzziconi, {\it {The $\Lambda$-BMS$_4$ charge algebra}},  {\em JHEP} {\bf 10} (2020) 205, [\href{http://arxiv.org/abs/2004.10769}{{\tt arXiv:2004.10769}}].

\bibitem{Geiller:2022vto}
M.~Geiller and C.~Zwikel, {\it {The partial Bondi gauge: Further enlarging the asymptotic structure of gravity}},  {\em SciPost Phys.} {\bf 13} (2022) 108, [\href{http://arxiv.org/abs/2205.11401}{{\tt arXiv:2205.11401}}].

\bibitem{Hartong:2025jpp}
J.~Hartong, E.~Have, V.~Nenmeli, and G.~Oling, {\it {Boundary Energy-Momentum Tensors for Asymptotically Flat Spacetimes}},  \href{http://arxiv.org/abs/2505.05432}{{\tt arXiv:2505.05432}}.

\bibitem{Fiorucci:2025twa}
A.~Fiorucci, S.~Pekar, P.~Marios~Petropoulos, and M.~Vilatte, {\it {Carrollian-holographic Derivation of BMS Flux-balance Laws}},  \href{http://arxiv.org/abs/2505.00077}{{\tt arXiv:2505.00077}}.

\bibitem{Maldacena:2000hw}
J.~M. Maldacena and H.~Ooguri, {\it {Strings in AdS(3) and SL(2,R) WZW model 1.: The Spectrum}},  {\em J. Math. Phys.} {\bf 42} (2001) 2929--2960, [\href{http://arxiv.org/abs/hep-th/0001053}{{\tt hep-th/0001053}}].

\bibitem{Adamo:2014yya}
T.~Adamo, E.~Casali, and D.~Skinner, {\it {Perturbative gravity at null infinity}},  {\em Class. Quant. Grav.} {\bf 31} (2014), no.~22 225008, [\href{http://arxiv.org/abs/1405.5122}{{\tt arXiv:1405.5122}}].

\bibitem{Adamo:2015fwa}
T.~Adamo and E.~Casali, {\it {Perturbative gauge theory at null infinity}},  {\em Phys. Rev. D} {\bf 91} (2015), no.~12 125022, [\href{http://arxiv.org/abs/1504.02304}{{\tt arXiv:1504.02304}}].

\bibitem{Mizera:2017cqs}
S.~Mizera, {\it {Combinatorics and Topology of Kawai-Lewellen-Tye Relations}},  {\em JHEP} {\bf 08} (2017) 097, [\href{http://arxiv.org/abs/1706.08527}{{\tt arXiv:1706.08527}}].

\bibitem{Mizera:2019gea}
S.~Mizera, {\em {Aspects of Scattering Amplitudes and Moduli Space Localization}}.
\newblock PhD thesis, Princeton, Inst. Advanced Study, 2020.
\newblock \href{http://arxiv.org/abs/1906.02099}{{\tt arXiv:1906.02099}}.

\bibitem{Mizera:2019blq}
S.~Mizera, {\it {Kinematic Jacobi Identity is a Residue Theorem: Geometry of Color-Kinematics Duality for Gauge and Gravity Amplitudes}},  {\em Phys. Rev. Lett.} {\bf 124} (2020), no.~14 141601, [\href{http://arxiv.org/abs/1912.03397}{{\tt arXiv:1912.03397}}].

\end{thebibliography}\endgroup

\end{document}